\RequirePackage[2020-02-02]{latexrelease}
\documentclass[%
superscriptaddress,
 amsmath,amssymb,
PRFluids,
]{revtex4-2}

\usepackage[utf8]{inputenc}
\usepackage{comment}
\usepackage{booktabs}
\usepackage{ar}
\usepackage{tikz}
\usepackage{graphicx}
\usepackage[version=4]{mhchem}
\usepackage{siunitx}
\usepackage{longtable,tabularx}
\usepackage{cleveref}
\usepackage{supertabular}
\usepackage{amsmath}
\usepackage{algpseudocode}
\usepackage{algorithm}
\usepackage[caption=false]{subfig}
\usepackage{relsize}
\usepackage{float}
\usepackage{bbding}
\usepackage{bm}
\usepackage{mathrsfs}
\usepackage{comment}
\graphicspath{ {./Figures/} }
\newcommand{\RN}[1]{%
  \textup{\uppercase\expandafter{\romannumeral#1}}%
}

\begin{document}
\title{Experimental mitigation of large-amplitude transverse gusts via closed-loop pitch control}
\author{Girguis Sedky}
 \affiliation{Department of Aerospace Engineering, University of Maryland, College Park, MD, 20742}
\author{Antonios Gementzopoulos}
 \affiliation{Department of Aerospace Engineering, University of Maryland, College Park, MD, 20742}
\author{Francis D. Lagor}
 \affiliation{Department of Mechanical and Aerospace Engineering, University at Buffalo, The State University of New York, Buffalo, NY, 14260}
\author{Anya R. Jones}
 \affiliation{Department of Aerospace Engineering, University of Maryland, College Park, MD, 20742}

\begin{abstract}

Air vehicles of all scales are susceptible to large-amplitude gusts that may lead to
vehicle damage and instability. Therefore, effective physics-based control strategies and
an understanding of the dominant unsteady flow physics underpinning gust encounters are
desired to improve safety and reliability of flight in these conditions. To this end, this paper
develops and experimentally demonstrates a proportional output-feedback lift regulation
strategy based on the classical unsteady aerodynamic theories of Wagner and K\"ussner for
wings encountering large-amplitude transverse gusts without a \textit{a priori} knowledge of gust
strength or onset time. The tested vertical gust velocities ranged between 25
the freestream speed. This strategy is found to successfully generalize to gusts of different
strengths and directions, as well as wings at pre- and post-stall angles of attack. In addition,
this paper applies dynamical systems analysis to Wagner’s aerodynamic model to reveal the
effect of the pitch-axis location, pitch input, and closed-loop feedback gains on the stability
and robustness of the system along with the flow physics responsible. The real-time output
feedback control strategy is experimentally tested in a water tow tank facility equipped with
a transverse gust generator. Time-resolved force and flow-field measurements are used to
discover the salient flow physics during these encounters and illustrate how closed-loop
actuation mitigates their lift transients.
\end{abstract}

\maketitle

\section{Introduction}
Air vehicles of varying scales respond to gust encounters~\cite{moulin2007gust,pines2006challenges}. For example, large-scale air vehicles, such as commercial aircraft, can be susceptible to gusts induced by turbulent flow structures in the atmospheric boundary layer~\cite{boettcher2003statistics,brasseur2001development} and stormy weather conditions. Strong gusts also lead to difficulties in takeoff and landing in the presence of strong cross winds or wakes shed from aircraft-carrier bluff-body structures~\cite{federal2011airplane,wilkinson1998modelling}. Even more vulnerable are smaller vehicles such as Uncrewed Air Vehicles (UAVs) or micro-air vehicles (MAVs) which can be susceptible to gusts arising from unsteady flows present in complex terrains and urban environments~\cite{zarovy2010experimental}. A small air vehicle's vulnerability is due to its low inertia which is easily overcome by small flow perturbations and its operation in a low Reynolds number regime which makes the vehicle more susceptible to flow separation. This work is primarily concerned with gusts of large amplitudes that can induce large aerodynamic load transients and immediate flow separation over the wing's surface. The possible dire consequences of large-amplitude gusts have spurred a large number of studies to understand the unsteady aerodynamics of wing-gust encounters~\cite{von1938airfoil,viswanath2010effect,Corkery2018} and to develop active and passive means for gust alleviation~\cite{cook2013robust,oduyela2014gust,bhatia2014stabilization}. The work presented here expands on the previous works by utilizing closed-loop pitch control to alleviate large-amplitude transverse gusts and examining the flow physics in detail using time-resolved force and flowfield measurements. 

A variety of methods currently exist for aircraft stability and control. Current fixed-wing aircraft rely on actuated control surfaces, such as ailerons, rudders, and elevators, to provide control~\cite{FAA2016Pilot}. Recent work experimentally demonstrated the effectiveness of using control surfaces in mitigating strong gusts. Herrmann et al. utilized trailing-edge flaps to alleviate vortex gusts and reported a 63.9\% reduction in the lift standard deviation during quasi-random gust encounters using a combined PI feedback/model-based feedforward architecture. Frederick et al.~\cite{Frederick2010gust} studied the use of a small, rapidly actuated, actively controlled trailing-edge flap (4\% chord) to alleviate the unsteady loads due to atmospheric turbulence and reported 79\% load reduction with the use of a proportional-integral-derivative (PID) controller. They also remarked that, theoretically, proportional control was sufficient in mitigating the loads, but experimentally was ineffective at controlling the unsteady loading with an upstream disturbance present. The work presented here builds on the study of Frederick et al.~\cite{Frederick2010gust} by providing experimental evidence of the ability of a proportional controller to mitigate upstream disturbances. The physical mechanisms underlying the success of this controller, mentioned briefly in their study, are also discussed in the present work. Another very prominent method of controlling aerodynamic loads on aircraft is the use of active flow control. Rather than using movable surfaces, an aircraft may use localized jet or plasma actuators that control the aircraft by forcing flow attachment or separation. Kersten et al.~\cite{kerstens2011closed} demonstrated the ability of variable-pressure pulsed-blowing actuation to mitigate longitudinal gusts. They noted that active flow control methods are effective for mitigating low-frequency disturbances but may experience a bandwidth limitation that may limit their effectiveness in mitigating higher-frequency disturbances. 

A control method seen less in engineered vehicles but ubiquitous in biological fliers is full wing and body maneuvers. Rather than moving or deflecting a control surface which comprises a small portion of the total wing planform area, biological flyers flap and rotate their entire wing. In fact, Vance et al.~\cite{vance2013kinematic} observed that stalk-eye flies used a combination of asymmetric flapping and wing pitching in response to gust perturbations. Andreu Angulo and Babinsky~\cite{andreu2022mitigation,andreu2022controlling} as well as Sedky et al.~\cite{sedky2022physics} conducted studies that utilized full wing open-loop pitch motions to mitigate large-amplitude transverse gusts. These studies demonstrated the ability of pitch maneuvers based on classical unsteady aerodynamic theory to effectively mitigate the lift transients from large-amplitude transverse gust encounters.  While these models were suitable for mitigating the lift transients experienced by the wing during the gust encounter, they poorly captured the lift response of the wing after exiting the gust for large angles of attack. As a result, Andreu Angulo and Babinsky~\cite{andreu2022controlling} found that these maneuvers exacerbated the lift transients in the gust recovery region for large angles-of-attack cases. The work presented here uses real-time lift measurements and a closed-loop feedback controller to compensate for the deficiency of these models in this regime. The present work used full-wing actuation to mitigate large-amplitude gusts for two reasons: (1) actuating the entire wing maximizes the forcing power of the control input as opposed to only actuating a small control surface, and (2) kinematic actuation provides instantaneous impact on the forces due to added-mass and instantaneous circulatory forces, and thus it does not suffer from aerodynamic bandwidth limitations.

This paper contributes to the study of lift regulation in gust encounters by designing and experimentally implementing a closed-loop output feedback controller using wing pitching as an input. The closed-loop controller is tested in a variety of conditions, including various gust strengths, starting angles of attack $\alpha^\text{ref.}$, and gust directions. Gust strength is parametrized using the gust ratio 
\begin{equation}
\text{GR}=\frac{v}{U_\infty},
\end{equation}
where $v$ is the maximum vertical velocity of the gust and $U_\infty$ is the freestream velocity. Time-resolved force and flowfield measurements are acquired to gain an understanding of the flow physics during the encounters and how they affect closed-loop mitigation. The contributions of this work are (1) the experimental demonstration of proportional feedback control for lift transient mitigation during wing-gust encounters; (2) a dynamical systems treatment of the classical aerodynamic model of Wagner that reveals how pitch-axis location, input, and control gain affect the stability and robustness of the system along with the flow physics responsible; and (3) discovery and understanding of the dominant flow physics behind lift transients during wing-gust encounters with and without control, underscoring the physical mechanisms with which closed-loop control mitigates lift.

\section{Dynamical systems analysis and control design}
\label{Ch:3}
Consider the lift mitigation problem of a wing encountering a transverse gust with pitching input based on real-time lift measurements. This section explores a principled approach to controller design based on linear systems and classical unsteady aerodynamic theory. In addition, the impact of control design choices, such as pitch-axis location, pitch input, and control gain, on the stability and robustness of the system along with the flow physics responsible are underscored.

\subsection{Closed-loop control design for gust rejection}
\begin{figure}[h]
     \centering
         \includegraphics[width=0.75\linewidth]{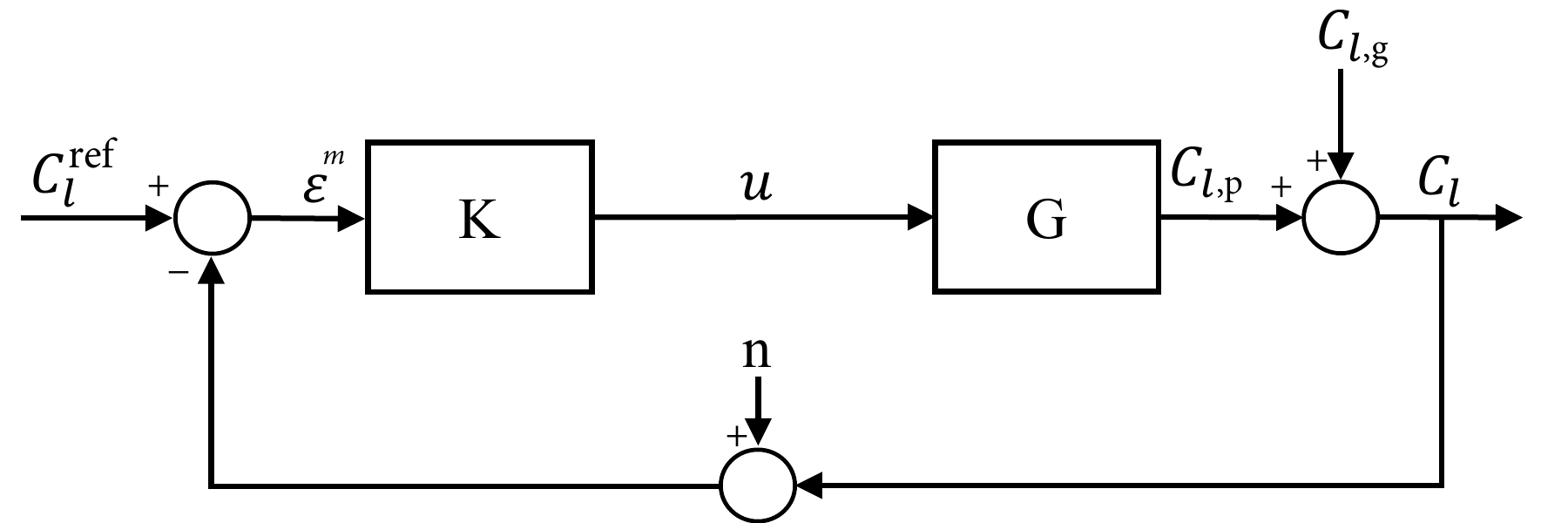}
         \caption{Closed-loop control diagram}  
          \label{fig:ControlDiagram}       
\end{figure}

The gust mitigation strategy explored in this work relies on linearizing the aerodynamics of wing-gust encounters, an assumption that was shown to hold well. For instance, Biler et al.~\cite{biler2018experimental} experimentally demonstrated the ability of the linear K\"ussner model to accurately predict the lift coefficients peaks during wing-gust encounters that exhibit fully separated flows and strong leading-edge vortex development. Andreu-Angulo et al.~\cite{andreu2021unsteady} and Sedky et al.~\cite{sedky2022physics} expanded this demonstration to pitching wings encountering transverse gusts. This assumption allows for (1) the computation of the total lift by linearly superimposing the impact of pitching alone using Wagner's aerodynamic model and the impact of the gust alone using K\"ussner's aerodynamic model, and (2) the utilization of many powerful tools from linear robust control theory designed to assess and mitigate the effect of disturbances and noise in the system. Figure~\ref{fig:ControlDiagram} shows the output feedback control framework used to design a controller that will regulate the lift of a wing about a desired value, $C_l^\text{ref}$. In this framework, the controller $K$ computes the control input $u$ based on the current value of the error, $\varepsilon=C_l^\text{ref}-C_l$. The calculated control input is applied to plant $G$ which governs the wing's aerodynamic lift response to the applied control pitch input $C_{l,\text{p}}$. The gust is assumed to produce a linearly additive lift disturbance $C_{l,\text{g}}$. The total $C_l$ measurement, based on the sum of the plant's lift response and the unknown gust disturbance, is fed back to the controller and the process repeats. Measured lift can be susceptible to noise, which is also added to $C_l$ before it is fed back to the controller. Note that the distinction between the true error $\varepsilon$ and the measured error $\varepsilon^m$ in figure~\ref{fig:ControlDiagram} is that $\varepsilon^m$ is corrupted by sensor noise. The measured error value is what is fed back to the controller, but the minimization of the true error is the control law's objective.

\subsection{The derivation of the plant $G$}
\begin{figure}[h]
     \centering
         \includegraphics[width=0.4\linewidth]{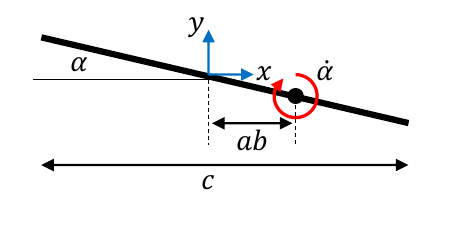}
         \caption{Sketch of a wing pitching about a point $ab$ away from the midchord}  
          \label{fig:WagnerParametersSketch}       
\end{figure}

The plant $G$, shown in figure~\ref{fig:ControlDiagram}, represents a transfer function with a pitch input and a lift output. To obtain the plant's transfer function, the unsteady aerodynamic model of choice can be transformed from the time domain to the frequency domain. Brunton et al.~\cite{brunton2013empirical} Laplace transformed Theodorsen's aerodynamic model into the complex frequency domain, but the same result can be obtained starting with Wagner's unsteady aerodynamic model. The version of the Wagner model used here is for a pitching wing but plunge and surge inputs may also be added. For a pitching wing, shown in figure~\ref{fig:WagnerParametersSketch}, with a semi-chord (half-chord) length $b$, the coefficient of lift $C_l$ in the time domain can be expressed as

\begin{equation}
C_l=\overbrace{\vphantom{\int}\pi\left( \dot{\alpha} - a \ddot{\alpha} \right)}^\text{added-mass}  + \overbrace{2\pi \int_{0}^{s_c} \frac{\text{d}\phi(\sigma)}{\text{d}\sigma}\left(\alpha(s_c-\sigma)+ \left(\frac{1}{2}-a\right)\dot{\alpha}(s_c-\sigma)\right) \,\text{d}\sigma}^\text{circulation},
\label{eq:TimeDomain}
\end{equation}

\noindent where $\phi$ is Wagner's potential function and $s_c$ is the displacement of the wing in semi-chords. Note that $s_c$ is also the non-dimensional convective time defined by semi-chords, $s_c=tU_\infty/b$. The pitching axis location of the wing is defined in semi-chords by the variable $a$ which is zero at the midchord, 1 at the trailing edge and -1 at the leading edge. All time derivatives are non-dimensional in equation~\ref{eq:TimeDomain}, i.e., $\dot{\alpha}=d\alpha/ds_c$ and $\ddot{\alpha}=d^2\alpha/ds_c^2$. The two main components in equation~\ref{eq:TimeDomain} are added mass and circulatory contributions. Added-mass force is classically described as the force needed to accelerate the fluid surrounding the body when the body accelerates~\citep{darwin1953note} and the circulatory contribution is the force due to growth and advection of circulation in the flow. The lift response to an angle-of-attack input can be obtained by taking the Laplace transform of equation~\ref{eq:TimeDomain} with respect to the angle of attack,
\begin{equation}
\mathcal{L}\{C_l\}=\pi \left(s\mathcal{L}\{\alpha\} - as^2\mathcal{L}\{\alpha\} \right)+2\pi\left[\mathcal{L}\left\{\frac{\text{d}\phi}{\text{d}s_c}\right\}  \left(\mathcal{L}\{\alpha\}+ \left(\frac{1}{2}- a \right)s\mathcal{L}\{\alpha\}\right)  \right],
\end{equation}

\noindent where $s$ is Laplace's variable non-dimensionalized by $b/U$. The Laplace variable $s$ can span the entire complex domain, but if restricted to the imaginary axis, then $s=ik_f$, where $k_f=b\omega/U$ is the reduced frequency. The Laplace transform of $\alpha$, $\mathcal{L}\{\alpha\}$, can be taken as a common factor and the Wagner potential function can be simplified, yielding 
\begin{equation}
\mathcal{L}\{C_l\}=\left(\pi s - \pi a s^2 +2\pi\left[s\phi(s)\left(1+\left(\frac{1}{2}-a\right)s\right)  \right]\right)\mathcal{L}\{\alpha\}.
\end{equation}

\noindent Garrick~\cite{garrick1939some} demonstrated the equivalency of Wagner's and Theodorsen's functions, which, if allowed to span the entire complex plane, can be expressed as
\begin{equation}
C(s)=s\phi(s).
\end{equation}
Thus, the expression can be further simplified to
\begin{equation}
\mathcal{L}\{C_l\}=G_\alpha \mathcal{L}\{\alpha\}=\left(\pi s - \pi a s^2 +2\pi\left[C(s)\left(1+\left(\frac{1}{2}-a\right)s\right)  \right]\right)\mathcal{L}\{\alpha\}.
\label{eq:Output1}
\end{equation}

The above expression yields a $\mathcal{L}\{C_l\}$ output for a $\mathcal{L}\{\alpha\}$ input, but since $\mathcal{L}\{\alpha\}=\mathcal{L}\{\dot{\alpha}\}/s=\mathcal{L}\{\ddot{\alpha}\}/s^2$, equation~\ref{eq:Output1} can be equivalently expressed as 
\begin{equation}
\mathcal{L}\{C_l\}=G_{\dot{\alpha}} \mathcal{L}\{\dot{\alpha}\}=\left(\pi - \pi a s +\frac{2\pi}{s}\left[C(s)\left(1+\left(\frac{1}{2}-a\right)s\right)  \right]\right)\mathcal{L}\{\dot{\alpha}\},
\label{eq:Output2}
\end{equation}
\noindent and
\begin{equation}
\mathcal{L}\{C_l\}=G_{\ddot{\alpha}} \mathcal{L}\{\ddot{\alpha}\}=\left(\frac{\pi}{s} - \pi a  +\frac{2\pi}{s^2}\left[C(s)\left(1+\left(\frac{1}{2} - a\right)s\right)  \right]\right)\mathcal{L}\{\ddot{\alpha}\}.
\label{eq:Output3}
\end{equation}
 
\noindent where Theodorsen's transfer function using the R.T. Jones approximation given by Brunton and Rowley~\cite{brunton2013empirical} is
\begin{equation}
C(s)=\frac{0.5 s^2 + 0.2808 s + 0.01365}{s^2 + 0.3455 + 0.01365}.
\label{eq:TheoApprox}
\end{equation}

Equations~\ref{eq:Output1},~\ref{eq:Output2}, and~\ref{eq:Output3} express the general relationship between the inputs $\alpha$, $\dot{\alpha}$, $\ddot{\alpha}$ to the output $C_l$ in the complex frequency domain. These relations can be used to understand the behavior and stability of the output $C_l$ in response to the input signals. In the following sections, these relations are utilized to understand the impact of the pitch axis, input choice, and closed-loop input gain on the lift experienced by the wing based on proportional output feedback control.

\subsection{Effect of pitch-axis location} 
\label{Sec:PitchAxis}
\begin{figure}[h]
\centering
\subfloat[0.4\textwidth][]{\includegraphics[width=0.4\textwidth]{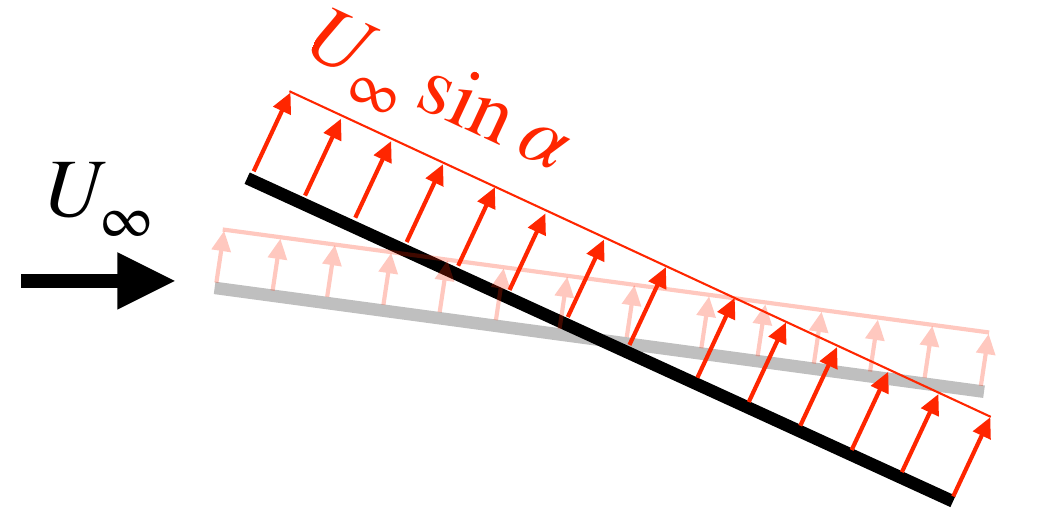}\label{fig:Component_1}}\hfill
\subfloat[0.4\textwidth][]{\includegraphics[width=0.4\textwidth]{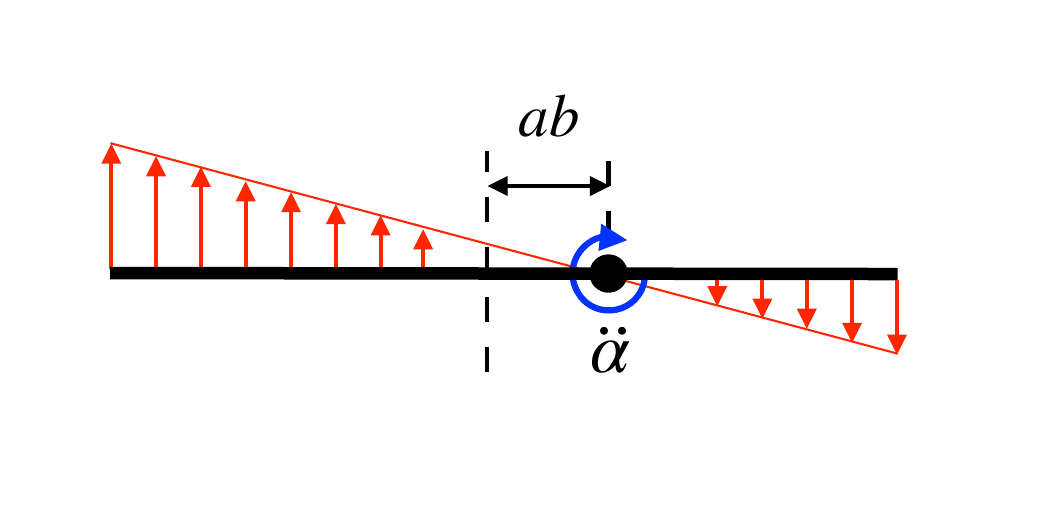}\label{fig:Component_2}}\hfill
\caption{Illustration of added-mass components ${\romannumeral 1}$ (a) and ${\romannumeral 2}$ (b) in eq~\ref{eq:TimeDomain} for a pitching-up wing}\label{fig:AddedMassComps}
\end{figure}

To utilize pitching to mitigate the aerodynamic loads of the gust, the pitch axis must be chosen carefully. The choice of the pitch axis $a$ changes the contributions of the added mass term $\pi a \ddot{\alpha}$ and the rotational circulation term $-(1/2 - a)\dot{\alpha}$ in equation~\ref{eq:TimeDomain}. To understand the impact of pitch location on the dynamics of the system, this section focuses on the added-mass contribution to the forces. This special focus is justified because this contribution dominates the aerodynamic forces during aggressive maneuvers. Figure~\ref{fig:AddedMassComps} visually illustrates the sources of the added-mass components in equation~\ref{eq:TimeDomain}. The pitch rate $\dot{\alpha}$ and pitch acceleration $\ddot{\alpha}$ are positive in the clockwise direction (pitch up motions). When a wing pitches, the freestream velocity component normal to the wing changes, as shown in figure~\ref{fig:Component_1}. Thus, in the wing's frame of reference, an acceleration in the freestream velocity component normal to the chord is observed. This component of added mass is invariant with respect to the pitch location because the same variation in the freesteam velocity component normal to the chord is observed irrespective of where the wing is pitching.

\noindent When the wing rotates with a non-zero pitch acceleration, $\ddot{\alpha} \neq 0$, it displaces an additional volume of fluid at a changing rate, as shown by the velocity distribution in figure~\ref{fig:Component_2}. When $a>0$ (pitching aft the midchord), the pitching wing displaces more fluid upstream than downstream of the pitching point. If $\ddot{\alpha}>0$, the net rate of change of the fluid displacement is positive. To create this net rate of change in the fluid's displacement, the wing imparts a positive upwards force on the fluid. The fluid reacts to the wing by imparting a downward force on it. If the aim of the pitch up motion is increasing lift, this component of added mass may reduce it instead which can lead to an unstable feedback loop. On the other hand, if $a<0$ (pitching before the midchord), the pitching wing displaces more fluid downstream than upstream of the pitching point. If $\ddot{\alpha}>0$, the net rate of change of the fluid displacement is negative. To create this net rate of change in the fluid's displacement, the wing imparts a negative downward force on the fluid. The fluid reacts to the wing by imparting an upward force on it. If the aim of the pitch up motion is to increase lift, this component of added mass acts to increase it.

The lift response to the input in the frequency domain can be studied for different pitch-axis locations using equation~\ref{eq:Output3}. In this analysis, a proportional closed-loop control law based on instantaneous coefficient of lift measurements is chosen,
\begin{equation}
\ddot{\alpha}=-k C_l,
\label{eq:ControlLaw}
\end{equation}
 
\noindent where $k$ is the control gain. Even though a pitch acceleration, $\ddot{\alpha}$, input is utilized to analyze the stability of the system based on pitching location, a study on the impact of control input is provided in section~\ref{Sec:InputImpact}. In this case, the closed-loop transfer function is
\begin{equation}
L_\text{closed}=\frac{kG_{\ddot{\alpha}}}{1 + kG_{\ddot{\alpha}}}. 
\end{equation}

 \begin{figure}[h]
     \centering
         \includegraphics[width=0.5\linewidth]{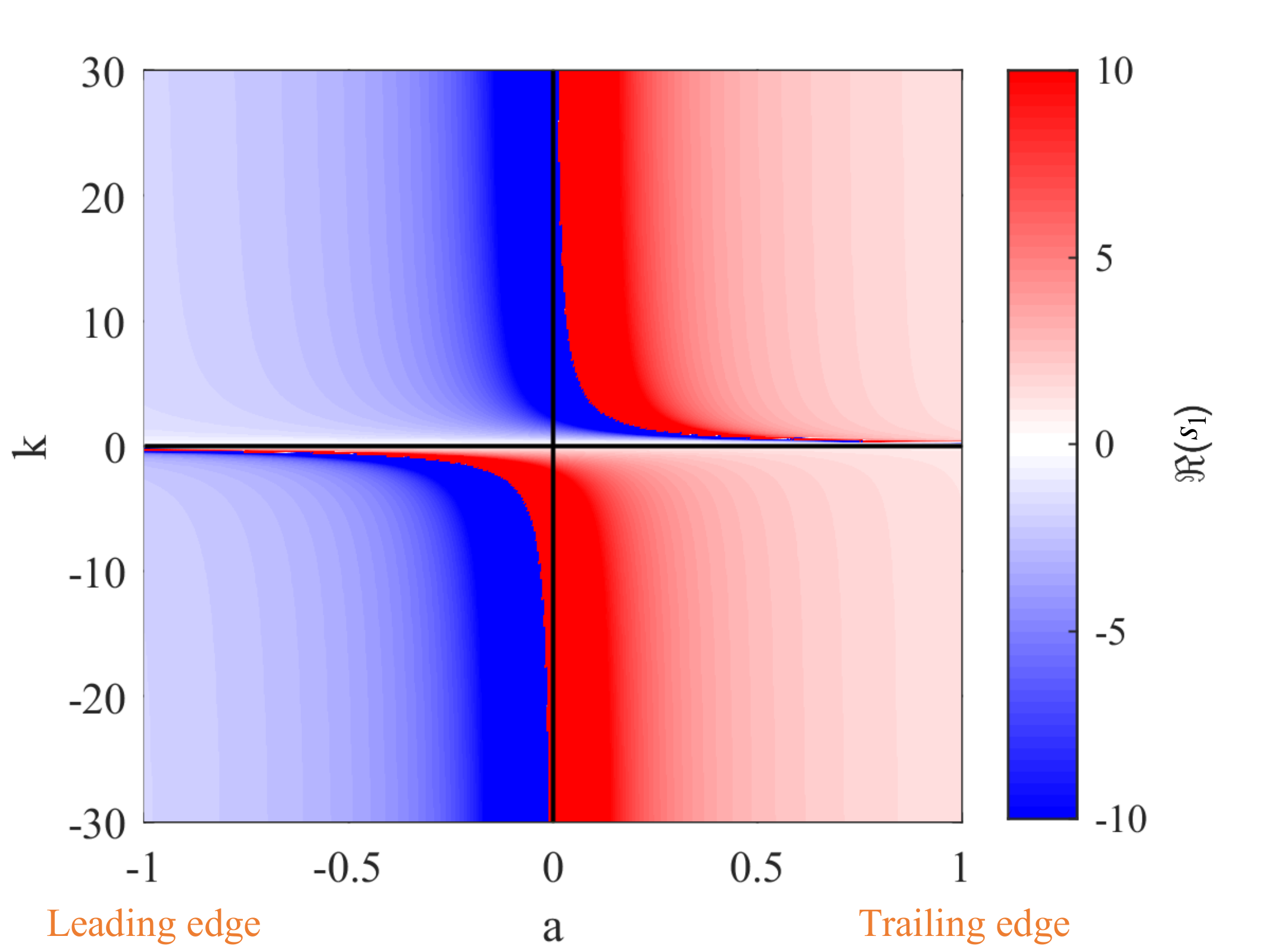}
         \caption{The influence of the control gain $k$ and the pitch axis $a$ on the location of the real part of $\Re(s_1)$ pole.}  
          \label{fig:PolesContour_PitchLocation_Gain_alphadotdot}       
\end{figure}

Stability of the system can be studied by observing the location of its poles, which are the roots of its characteristic equation, in the complex plane. The minimal representation of this system (with pole-zero cancellation) has four poles. Only the first pole of the system $s_1$ is significantly affected by the change in pitching axis and gain. Thus, variation in the real part of the first pole, $\Re{(s_1)}$, as a function of the gain and the pitching axis can be studied to understand the impact of the pitch location on stability of the system. The system is exponentially stable if $\Re{(s_1)}<0$ and unstable if $\Re{(s_1)}>0$. Figure~\ref{fig:PolesContour_PitchLocation_Gain_alphadotdot} shows a colored contour of how the first pole varies as a function of the gain chosen, $k$, and the pitch axis $a$. The blue shading indicates stability of the $s_1$ pole with $\Re(s_1)<0$. Pitching forward of the midchord tends to be aerodynamically stable for positive and negative gains. Pitching aft of the midchord tends to be unstable for positive and negative gains. Pitching about the midchord is only stable for $k>0$. The change in stability as the pitching-axis location passes the midchord is due to the added-mass component discussed in this section. Moving the pitching point after the midchord creates added-mass forces that oppose the lift control objective.

\subsection{Effect of pitch input choice}
\label{Sec:InputImpact}
In this section, the pitching axis is fixed to the wing's midchord, $a=0$, and the control input is varied to understand the impact of control input choice on the system's performance. Setting $a=0$, the transfer functions in equations~\ref{eq:Output1},~\ref{eq:Output2}, and~\ref{eq:Output3} based on the R.T. Jones approximation of Theodorsen's function (equation~\ref{eq:TheoApprox}) can be simplified to 
\begin{equation}
G_{\alpha}=\frac{4.71s^3+5.11s^2+1.85s+0.09}{s^2+0.35s+0.01},
\label{eq:Output1_simp}
\end{equation}
\begin{equation}
G_{\dot{\alpha}}=\frac{4.71s^3+5.11s^2+1.85s+0.09}{s(s^2+0.35s+0.01)},
\label{eq:Output2_simp}
\end{equation}
\begin{equation}
G_{\ddot{\alpha}}=\frac{4.71s^3+5.11s^2+1.85s+0.09}{s^2(s^2+0.35s+0.01)}.
\label{eq:Output3_simp}
\end{equation}

The transfer functions $G_\alpha$, $G_{\dot{\alpha}}$, and $G_{\ddot{\alpha}}$ govern the response of the coefficient of lift $C_l$ to pitch $\alpha$, pitch rate $\dot{\alpha}$, and pitch acceleration ${\ddot{\alpha}}$, respectively. Important response features of the system and how they vary with input choice can be obtained from understanding the structure of these transfer functions. Note that the open-loop transfer function $G_\alpha$, does not contain a pole at the origin. According to the Internal Model Principal~\cite{skogestad2005multivariable}, the addition of integral control is required for this control input to ensure accurate tracking of a constant reference lift signal. In addition, the behavior of this transfer function for high-frequency input may be studied by taking the limit of $G_\alpha$ as the frequency of the input $\omega$ approaches infinity. Since the degree of the numerator of the transfer function is higher than the denominator,
\begin{equation}
\lim_{\omega\to\infty} |G_\alpha(\text{j}\omega)| = \infty,
\end{equation}

\noindent which indicates that input signals with a frequency approaching infinity will be amplified by an infinite amplitude ratio. Thus, high-frequency noise, which is common in practical applications, can be greatly amplified in the output. Lack of robustness to high-frequency noise makes the $\alpha$ input undesirable. 

$G_{\dot{\alpha}}$ has an instantaneous transfer between input and output, i.e., some element of the input signal (no matter how high its frequency) feeds through to the output without delay. For $G_{\dot{\alpha}}$, the amplitude ratio is
 
\begin{equation}
\lim_{\omega\to\infty} |G_{\dot{\alpha}}(\text{j}\omega)| = 4.71.
\end{equation}

\noindent Unlike $G_\alpha$, input signals with a frequency approaching infinity do not result in an infinite output response. However, they are not attenuated either, which makes this input choice also not robust to noise in the sensor measurements. Since $G_{\dot{\alpha}}$ has a pole at $s=0$, integral control is not required to achieve perfect tracking for constant output signals.

The limit of $G_{\ddot{\alpha}}$ as the frequency of the input $\omega$ approaches infinity is 
\begin{equation}
\lim_{\omega\to\infty} G_{\ddot{\alpha}} = 0.
\end{equation}

\noindent Unlike $G_\alpha$ and $G_{\dot{\alpha}}$, input signals with a frequency approaching infinity get attenuated. This makes the system under pitch acceleration input $\ddot{\alpha}$ more robust to sensor noise. Since the system also has poles at $s=0$, integral control is not required to achieve perfect tracking for constant output signals. Due to the tracking and sensor noise robustness properties of $G_{\ddot{\alpha}}$, commanding pitch acceleration is chosen in this work.

To understand the properties of the transfer functions presented from an aerodynamic perspective, the relation between lift and added mass in equation~\ref{eq:TimeDomain} can be examined with $a=0$. The dependence of the lift coefficient on the added-mass force term $\pi\dot{\alpha}$ is strictly algebraic. This means that lift will respond instantly to a pitch rate input without any lag. Physical systems exhibit a non-trivial lag between input and output, but the assumption of the infinite propagation velocity of information in incompressible flows allows for a response without any lag. Due to the feed-through added-mass term, any noise in the input signal shows up in the output signal. Note that the circulatory contribution to the force also has an instantaneous feed-through portion discussed by Taha and Rezaei\cite{taha2020high} that contributes to this behavior. Force and pressure sensors are prone to high frequency noise for many reasons, including flow turbulence, electromagnetic interference from nearby electronics, and structural vibrations. If an output feedback controller based on lift measurements is used, sensor measurement noise will be passed to the input signal which will in turn lead to additional high frequency lift response, which will then induce high frequency input signals and so on. If pitch input $\alpha$ were chosen for the control input, sensor noise would show up in the pitch profile instead. In this case, the added-mass component depends on the time derivative of the pitch profile $\text{d}\alpha/\text{d}t$. Taking a derivative of a signal amplifies the noise which leads to even higher feed-through noise. As a result, pitch actuation leads to a system less robust to noise. On the other hand, since $\dot{\alpha}=\int_{0}^{t} \ddot{\alpha} \text{d}t$, when pitch acceleration input $\ddot{\alpha}$ is used, it is integrated in time before passing to the output. Since integrating a noisy signal can smooth it, using $\ddot{\alpha}$ as the input is the choice that is most robust to high-frequency sensor noise.   

\subsection{Effect of gain choice}
\begin{figure}[h]
     \centering
         \includegraphics[width=0.4\linewidth]{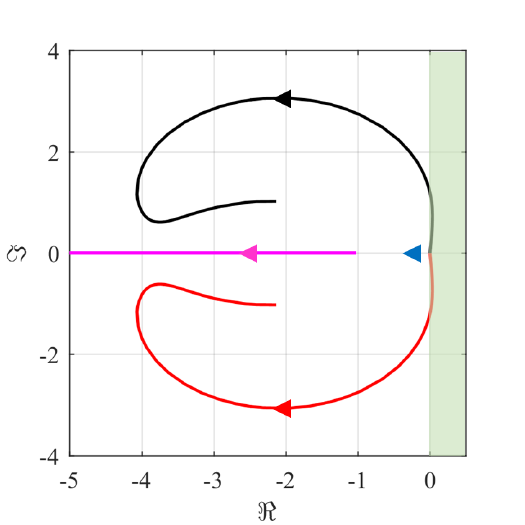}
         \caption{Root locus of the system.}  
          \label{fig:RootLocus}       
\end{figure}

This section studies the impact of the gain used in proportional control based on equation~\ref{eq:ControlLaw} on the aerodynamics of the system for a pitch acceleration input and midchord pitch location. To study the gain's influence on system stability, the root locus depicts how the poles of the system move in the complex plane as the gain $k$ varies from 0 to $\infty$. Figure~\ref{fig:RootLocus} shows how each of the four poles of the system vary in the complex plane. The arrows indicate how the poles move as the gain is increased. The poles must lie outside of the right half-plane, highlighted in green, to ensure stability of the closed-loop system. The system pole highlighted in blue remains in the same stable position in the left half-plane for all the gains. The pole highlighted in pink contains no imaginary component and starts at $\Re{(s)}=-1$ and moves to the left of the plane as $k$ increases. The remaining two poles, highlighted in red and black, are a complex conjugate pair. They start at the origin, move into the right-half plane as the gain increases, then move back into the left-half plane and converge to a pair with a real part $\Re{(s)} \approx -2$. Thus, the system is unstable for small gains and increasing the gain beyond a critical value stabilizes it.

\begin{figure}[h]
\centering
\subfloat[0.49\textwidth][$k=0.05$]{\includegraphics[width=0.49\textwidth]{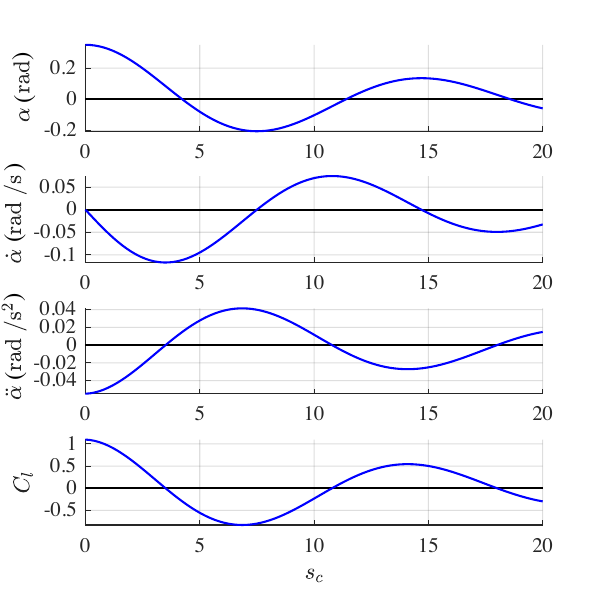}\label{fig:SmallGain}} \hfill
\subfloat[0.49\textwidth][$k=1.00$]{\includegraphics[width=0.49\textwidth]{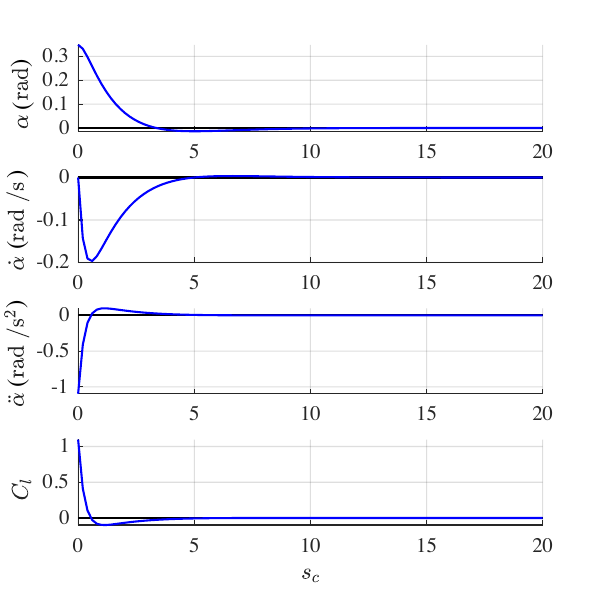}\label{fig:LargeGain}}\hfill
 \caption{Angle of attack $\alpha$, pitch rate $\dot{\alpha}$, and lift response $C_l$ for a controlled wing starting at an initial positive angle of attack $\alpha=20^{\circ}$ for two different proportional gains $k$.}
 \label{fig:WagnerStep}
\end{figure}

To understand why a small gain undermines the stability of the system while a larger gain improves stability, the dynamics of the aerodynamic model are simulated with an initial positive angle of attack $\alpha=20^{\circ}$ under closed-loop control. For a stable system, the lift coefficient converges to $C_l=0$, which corresponds to a steady-state angle of attack $\alpha=0^{\circ}$. Figure~\ref{fig:SmallGain} shows the system response for gain $k=0.05$ and figure~\ref{fig:LargeGain} shows the system response for gain $k=1$. The key to understanding the different behavior between the two cases is observing the angle of attack $\alpha$ at the point where $C_l$ crosses the $x$-axis. For $k=0.05$, $C_l$ and $\alpha$ cross the $x$-axis simultaneously. In other words, the control signal reverses only upon $\alpha$ passing the $x$-axis, and thus the wing only begins to decelerate after the wing passes its angle of attack goal $\alpha=0$. As a result, the wing never decelerates towards its goal, rather it oscillates about $\alpha=0$. 

Figure~\ref{fig:LargeGain} shows that for a large gain, $C_l$ crosses the $x$-axis at a positive angle of attack. At $s_c=0$, it begins to pitch down rapidly to $\alpha=0$. Due to its high negative pitch rate, the wing achieves a negative lift coefficient at a positive angle of attack. This causes a reversal in the control signal as the wing pitches towards $\alpha=0$, decelerating towards its goal. The lift reversal in this case is caused by the increase in the added-mass and circulatory lift contributions shown in equation~\ref{eq:TimeDomain} that are associated with high pitch rates $\dot{\alpha}$. Thus, choosing a high gain in the control law in equation~\ref{eq:ControlLaw} improves stability by activating unsteady flow physics that dampen the response of the system.

\subsection{Control design for disturbance and noise rejection}
While increasing the gain in the control law in equation~\ref{eq:ControlLaw} improves stability, it may make the system susceptible to high-frequency sensor noise. This section selects an appropriate gain for the controller based on a disturbance-noise rejection trade-off. Based on the block diagram in figure~\ref{fig:ControlDiagram}, the output lift coefficient free of sensor noise can be expressed as 

\begin{equation}
C_l=Gu + C_{l,\text{g}}=GK\varepsilon^m(s) + C_{l,\text{g}}.
\label{eq:ClosedLoopTFDeriv_1}
\end{equation}

\noindent The measured error signal between the measured and reference coefficients of lift is

\begin{equation}
\varepsilon^m(s)=C_l^\text{ref}-(C_l+n).
\label{eq:ClosedLoopTFDeriv_2}
\end{equation}

\noindent By substituting equation~\ref{eq:ClosedLoopTFDeriv_2} into equation~\ref{eq:ClosedLoopTFDeriv_1}, $C_l$ can be expressed as 

\begin{equation}
C_l=\frac{GK}{1+GK}C_l^\text{ref} - \frac{GK}{1+GK}n + \frac{1}{1+GK}C_{l,\text{g}}.
\label{eq:ClosedLoopTFDeriv_3}
\end{equation}

\noindent Let the open-loop transfer function be $L=GK$ which governs the $C_{l,\text{p}}$ response to the measured error signal $\varepsilon^m$. The controller should limit deviations of the true lift signal $C_l$ from the desired lift signal $C_l^\text{ref}$. The error value can be given in terms of $L$ as

\begin{equation}
\begin{split}
\varepsilon(s)=C_l^\text{ref}-C_l & = C_l^\text{ref}-\left( \frac{L}{1+L}C_l^\text{ref} - \frac{L}{1+L}n + \frac{1}{1+L}C_{l,\text{g}} \right) \\
 & =  \underbrace{\frac{1}{1+L}}_{S}C_l^\text{ref} + \underbrace{\frac{L}{1+L}}_T n - \underbrace{\frac{1}{1+L}}_SC_{l,\text{g}},
\end{split}
\label{eq:errorEq}
\end{equation}

\noindent where $S$ represents the sensitivity transfer function and $T$ represents the complementary sensitivity transfer function. Equation~\ref{eq:errorEq} shows how the error signal responds to the reference $C_l^\text{ref}$, noise $n$, and disturbance $C_{l,\text{g}}$ signals. More specifically, the frequency characteristics of the sensitivity transfer function $S$ determine how the error signal responds to variations in the reference and disturbance signals, while the complementary sensitivity transfer function $T$ determines how the error signal responds to variations in the noise signal. The `complementary' nature of these transfer functions,

\begin{equation}
S+T=\frac{1}{1+L}+\frac{L}{1+L}=1,
\label{eq:TF_Comp}
\end{equation}

\noindent implies that there is a trade-off between disturbance rejection, which requires $S$ to be small, and noise rejection, which requires $T$ to be small. It is important to note that the magnitude of each transfer function varies with frequency even though the relationship~\ref{eq:TF_Comp} holds at all frequencies. Thus, if the disturbance and noise frequencies are an order of magnitude or more apart, it is possible to choose a controller that will lead to a small sensitivity transfer function magnitude for the expected frequency range of the gust disturbance and a small complementary sensitivity transfer function magnitude for the expected frequency range of the sensor noise.
 
\begin{figure}[h]
     \centering
         \includegraphics[width=0.5\linewidth]{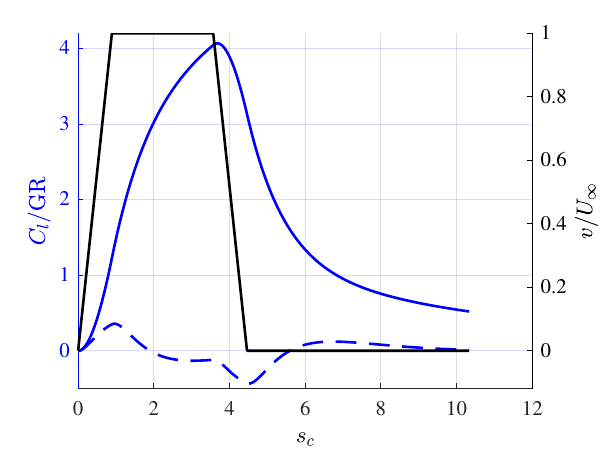}
         \caption{Normalized trapezoidal gust profile (solid black), the simulated normalized coefficient of lift based on K\"ussner's model (solid blue), and the simulated normalized coefficient of lift with closed-loop feedback control based on K\"ussner's and Wagner's models (dashed blue).}  
          \label{fig:Kussner}       
\end{figure}

In this paper, the controller is selected based on an expected frequency range for the gust. The reduced frequency of sinusoidal gusts, such as the canonical sine-squared gust, has been used in previous works for the purposes of kinematic gust modeling and control~\cite{Sedky2020lift,leung2018modeling}. For more complicated gust profiles, the gust disturbance should ideally be broken into individual frequency constituents to choose the controller. In addition, discontinuous profiles such as top-hat or trapezoidal gusts may contain high-frequency content due to abrupt transitions in their velocity profiles. For instance, a top-hat or a step gust will have an infinite range of frequencies due to the discontinuous jump in velocity. Figure~\ref{fig:Kussner} shows K\"ussner's model lift response for a trapezoidal gust encounter. Note that lift takes time to build up which limits the effectiveness of a high frequency gust forcing. Thus, in this work, the frequency content of the gust disturbance is based on the lift signal it produces rather than the gust profile directly. 

\begin{figure}[h]
     \centering
         \includegraphics[width=0.5\linewidth]{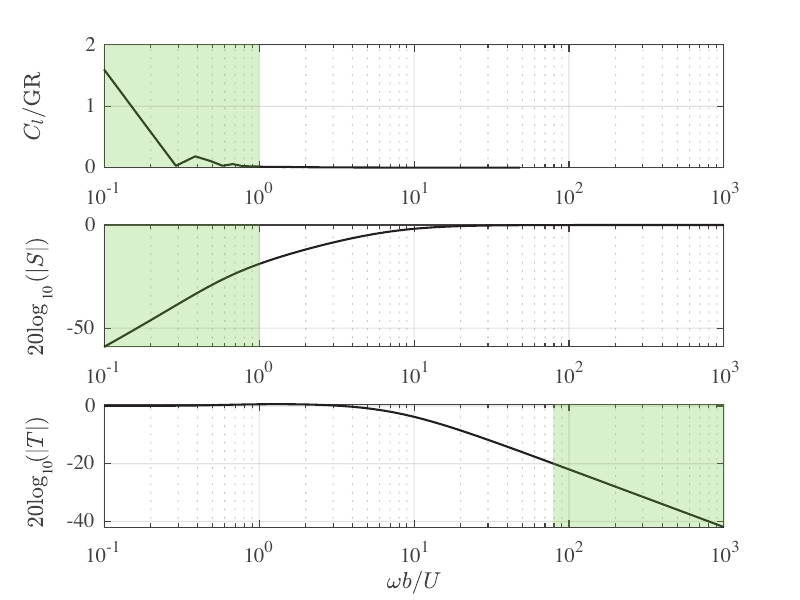}
         \caption{K\"ussner's lift response $C_L/\text{GR}$ in the frequency domain along with the amplitude ratios of the sensitivity transfer function $S$ and the complementary sensitivity transfer function $T$ at different reduced frequencies.}  
          \label{fig:DisturbanceSubPlot}       
\end{figure}

\noindent The unmitigated lift response to the gust according to K\"ussner's model shown in figure~\ref{fig:Kussner} was transformed to the frequency domain using a discrete Fourier transform. Figure~\ref{fig:DisturbanceSubPlot} shows this response along with the amplitude ratios for the sensitivity transfer function and the complementary sensitivity transfer function at different reduced frequencies for the chosen gain value $k=1.7$. The first subplot shows that as the reduced frequency increases, the frequency content in the lift response signal decreases. The sensitivity transfer function shown in the second subplot provides low gains in the low frequency range, indicating that low-frequency disturbances get attenuated the most. The green region highlighted in the first two subplots correspond to $20\log_{10}(|S|) \leq -20$, which is the frequency range where the controller can attenuate the gust by at least 90\%. It is evident from the top plot that this region encompasses almost all of the frequency content of the gust. The third plot shows the complementary transfer function's frequency response with the frequency range in which the controller is able to attenuate the noise by at least 90\% highlighted in green. This region corresponds to a reduced frequency $\omega b/U \geq 82$ or a dimensional frequency $f \geq 45$ Hz, which is a low frequency for common sensor noise due to electromagnetic interference. In addition, a moving average filter is applied to the sensor measurements in experiments to reduce the effect of lower frequency structural vibrations in the water tow tank's linear carriage and vertical control rods. Figure~\ref{fig:Kussner} shows the simulated lift signal for the wing encountering the gust with closed-loop feedback control based on K\"ussner's and Wagner's aerodynamic models. This represents the theoretical limit on the controller's efficacy which will be compared to the controller's actual performance in section~\ref{Ch:7}.

\subsection{Section summary}
This section describes the control design framework utilized in this work. Wagner's aerodynamic model is cast in the frequency domain and the aerodynamic transfer functions that govern the $C_l$ response to a pitch $\alpha$, pitch rate $\dot{\alpha}$, and pitch acceleration $\ddot{\alpha}$ are derived. The impact of pitch location, input level, and gain choice on the aerodynamic system are studied in detail. 

\begin{table}[h]
\centering
\caption{Effects of pitch location, input levels, and gain choice on closed-loop stability of proportional output-feedback control}
\label{tab:my-table}
\resizebox{\textwidth}{!}{%
\begin{tabular}{c||c|l|l}
\hline
               & Control choice  & \multicolumn{1}{c|}{Attribute} & \multicolumn{1}{c}{Reason}                                                               \\ \hline \hline
Pitch Location & $a \leq 0$      & 1 - Stable                         & 1 - Added-mass supports lift control goal                                                    \\ \cline{2-4} 
               & $a > 0$       & 1 - Unstable                       & 1 - Added-mass opposes lift control goal                                                     \\ \hline
Input          & $\dot{\alpha}$  & 1 - Track constant lift signal: \textcolor{green}{\checkmark}   & 1 - $L$ contains 1 pole                                                                      \\
               &                 & 2 - Robust to sensor noise: \textcolor{red}{\XSolidBrush}               & 2 - Sensor noise directly feeds through to lift due to added-mass and rotational circulation \\ \cline{2-4} 
               & $\alpha$        & 1 - Track constant lift signal: \textcolor{red}{\XSolidBrush}    & 1 - $L$ contains 0 poles                                                                     \\
               &                 & 2 - Robust to sensor noise: \textcolor{red}{\XSolidBrush}              & 2 - Sensor noise is amplified, $\dot{\alpha}=\text{d}\alpha/\text{d}t$                                    \\ \cline{2-4} 
               & $\ddot{\alpha}$ & 1 - Track constant lift signal: \textcolor{green}{\checkmark}   & 1 - $L$ contains 2 poles                                                                     \\
               &                 & 2 - Robust to sensor noise: \textcolor{green}{\checkmark}              & 2 - Sensor noise is attenuated, $\dot{\alpha}=\int_{0}^{t} \ddot{\alpha} \text{d}t$          \\ \hline
Gain           & Low             & 1 - Unstable or marginally stable  & 1 - Quasi-steady response results in an oscillatory behavior                                 \\ \cline{2-4} 
               & High            & 1 - Stable                         & 1 - Unsteady physics (added-mass and rotational circulation) dampen the system               \\ \hline
\end{tabular}%
}
\end{table}

Table~\ref{tab:my-table} summarizes the findings of the impact of pitch location, input level, and gain used in the proportional output-feedback control law on the stability and performance of the system. The final gain is chosen based on a disturbance and noise rejection trade-off. The frequency range of the lift disturbance signal expected from the trapezoidal gust profile is found using a discrete Fourier transform and a proportional gain is chosen such that it provides adequate disturbance and noise rejection. The sensitivity and complementary sensitivity transfer function frequency responses with the chosen gain indicate 90\% rejection rate for the expected disturbance and noise frequency ranges. 

\section{Experimental methodology}
\label{Ch:5}
This section describes the setup and methodology for force and flowfield data acquisition for wing-gust encounter experiments. In addition, it provides a description of the trapezoidal transverse gust generator used in these experiments as well as the implementation of the closed-loop control algorithm used in gust mitigation.

\subsection{Facility and test model}
\label{Sec:Facility and test model}
\begin{figure}[h]
\centering
  \includegraphics[width=\linewidth]{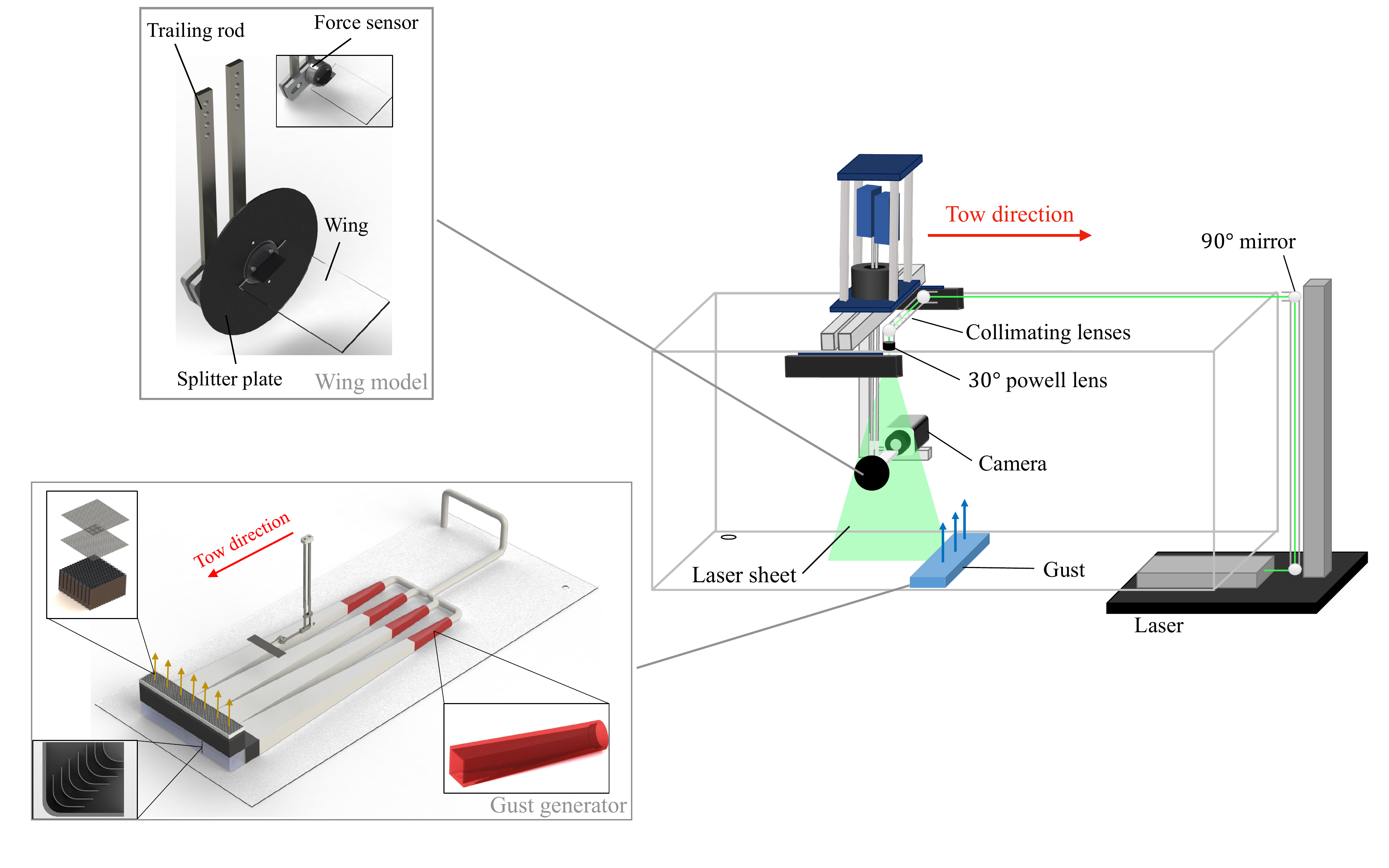}
\caption{Schematic of the experimental set-up}
\label{fig:PIV_Sketch}
\end{figure}

Experimental data collection was conducted in a free-surface water towing tank at the University of Maryland. Figure~\ref{fig:PIV_Sketch} shows the towing tank, the wing-test article, and the transverse gust generator. The wing was towed through the water until it encountered the transverse gust at the center of the tank. The wing-test article has a slider mechanism designed to pitch the wing about its midchord. Pitching actuation was accomplished by fixing the vertical height of the leading control rod and varying the vertical height of the trailing control rod. A half-span wing was used with a 1/16 in. stainless-steel splitter plate. The splitter plate isolated the half-wing from flow disturbances caused by the pitching mechanism and force balance. Experiments were conducted at Reynolds number $Re=10,000$ using a half-wing flat-plate model with a chord length $c = 0.0762 \ \text{m}$,  an effective aspect ratio \AR$ = 4$ based on symmetry about the splitter plate, and a thickness-to-chord ratio equal to 0.0417.

The gust generator shown in figure~\ref{fig:PIV_Sketch} provided the transverse gust flow field at the center of the water tow tank. The gust jet is shown with orange arrows. A variable speed 1.85 HP Hayward centrifugal pump drove the recirculating gust system. The gust ratio was varied by varying the speed of the pump. The generator created a gust with a trapezoidal velocity profile. More details regarding the design and characterization of the gust generator can be found in~\cite{sedky2022experimental}.

\subsection{Flowfield measurements}
\label{Sec:Force and torque measurements}
Planar Particle Image Velocimetry (PIV) was used to acquire spatially- and time-resolved measurements of the flow around the wing during wing-gust encounter experiments. The laser sheet and the camera of the PIV system were towed with the wing such that the wing remained in the camera's field of view. The PIV setup utilized a Litron high-speed Nd:YLF 532 nm laser with an acquisition rate of 700 Hz. The laser head was situated under the water tow tank, and the laser beam was redirected towards the area of interest via a series of $90^\circ$ mirrors. The beam was re-collimated using a series of converging and diverging lenses before arriving to a Powell lens that fanned the beam into a laser sheet. The laser sheet illuminated neutrally buoyant class IV soda lime spheres of 37 $\mu$m diameter. The images acquired were processed using multi-pass cross-correlation in LaVision's DaVis software. A $48 \times  48$ px square interrogation window was used for the first three correlation passes and a $24  \times  24$ px adaptive interrogation window was used for the following three passes. Interrogation regions were overlapped by 50\%. Post-processing was performed using a median universal outlier filter with a $5 \times 5$ filter region for the gust encounter cases. All force and flowfield measurements were ensemble averaged for each case. The final PIV vector spacing is 1.7 \% of the chord length. A built-in DaVis error tool based on the work of Wieneke~\cite{wieneke2015piv} was used to quantify the uncertainty in the PIV measurements. The maximum measurement uncertainty was determined to be 5.8\% within the center of the leading-edge vortex. When interpreting the flowfield measurements, it is also important to keep in mind the role of three-dimensional effects for the finite-aspect-ratio wing. The experimental work of Biler~\cite{BilerThesis} and the computational work of Badrya et al.~\cite{badrya2021} demonstrated that the flow topology remains consistent over most of the wing's span during the first half of the wing-gust encounter (i.e., until the lift peak), except near the wing's tip.

\subsection{Force measurements and implementation of closed-loop control}
\label{Sec:Implementation of closed-loop control}

\begin{figure}[h]
\centering
  \includegraphics[width=\linewidth]{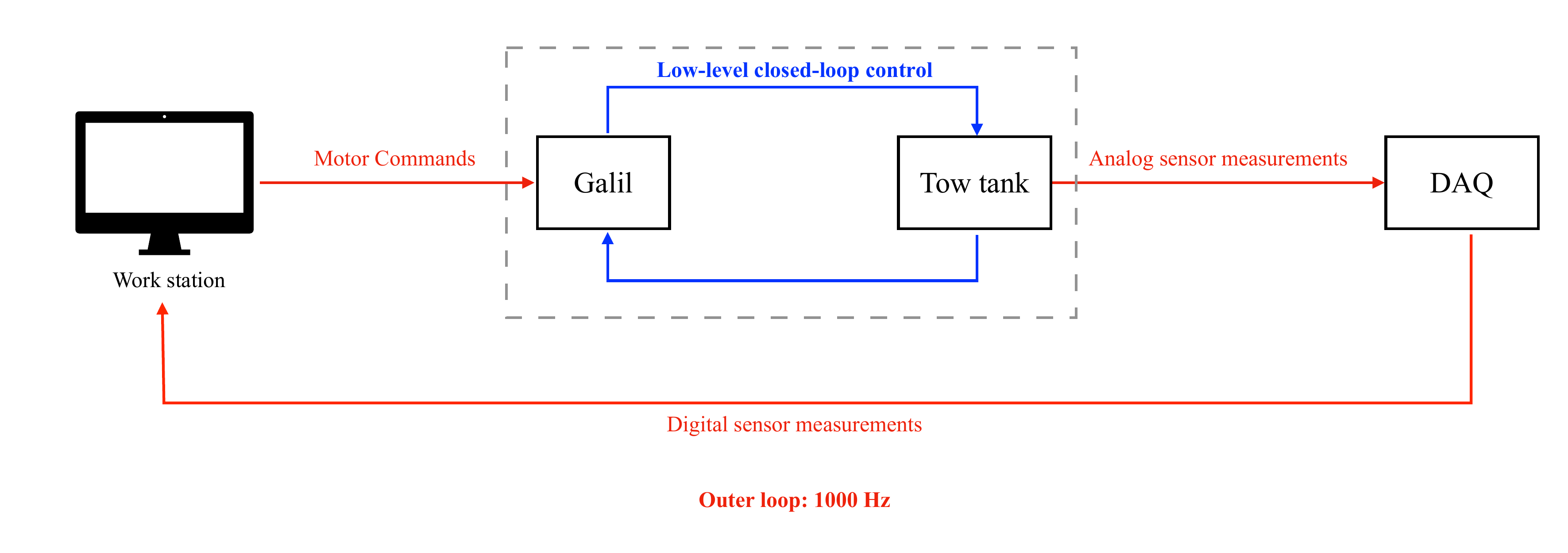}
\caption{Closed-loop control setup}
\label{fig:ClosedLoop_Sketch}
\end{figure}

Figure~\ref{fig:ClosedLoop_Sketch} shows a schematic of the hardware connections utilized in the closed-loop control experiments. Forces were acquired from a six degree-of-freedom ATI Mini-40 force sensor using a multi-functional NI USB6210 DAQ device. The uncertainty in the force measurements based on the lift coefficient peak at $\text{GR}=0.50$ is equal to $\pm$2.15\%. Since the force sensor rotates with the wing during the pitching motion, static angle-of-attack sweeps were conducted in the range of expected wing pitch angles prior to the closed-loop experiments. The tare forces and moments measured in these sweeps account for sensor bias, the weight of the test article, and the buoyancy force. During closed-loop experiments, the sensor was continuously tared based on the recorded tare measurements and the current wing orientation to obtain the aerodynamic loads. The outer loop in figure~\ref{fig:ClosedLoop_Sketch} implements the aerodynamic feedback controller. The control law was run on the computer of the water tow tank. Through Matlab, the computer acquired digital measurements in real-time from the DAQ and updated the motor actuation by sending out new reference motor commands. The reference commands were received by a low-level, high-precision PID motor controller made by Galil that ensures the tow-tank motors track the most recent reference command issued by the computer.

At time step $n$, the lift force $L$ was calculated based on the measured aerodynamic loads in the wing's body frame and the current angle of attack $\alpha^(n)$,

\begin{equation}
L^{(n)}=F_x^{(n)}\cos{\alpha^{(n)}}-F_y^{(n)}\sin{\alpha^{(n)}},
\end{equation}

\noindent where $F_x$ is the aerodynamic load component normal to the chord pointing upward, and $F_y$ is the aerodynamic load component parallel to the chord pointing towards the trailing edge. The coefficient of lift,
\begin{equation}
C_L^{(n)}=\frac{L^{(n)}}{(1/2) \rho U_\infty^2S},
\end{equation}

\noindent was obtained by dividing the lift force by the dynamic pressure, $1/2 \rho U_\infty^2$, and the wing's planform area $S$. The new pitch acceleration command for time step $n$ based on the chosen control law was calculated to be

\begin{equation}
\ddot{\alpha}^{(n)}=-k \left(C_L^{(n)} -C_L^\text{ref} \right).
\end{equation}

\noindent The motors were operated based on reference velocity commands supplied to the Galil controller, so the pitch accelerations were numerically integrated based on a simple Euler scheme to obtain the corresponding pitch rate $\dot{\alpha}$

\begin{equation}
\dot{\alpha}^{(n+1)}=\ddot{\alpha}^{(n)}\Delta t + \dot{\alpha}^{(n)}.
\end{equation}

\noindent The velocity command for the linear actuator controlling the trailing support rod shown in figure~\ref{fig:PIV_Sketch} was computed based on $\ddot{\alpha}^{(n)}$ and $\dot{\alpha}^{(n+1)}$ and sent to the motor controller.

\section{Results and discussion}
\label{Ch:7}
In this section, the closed-loop control experimental results are presented for lift tracking at angles of attack $\alpha^\text{ref.}=0^{\circ}$, $5^{\circ}$, $10^{\circ}$, and $15^{\circ}$, and gust ratios $\text{GR}=0.25$, $0.50$, and $0.71$. The experimental results include transverse upwards and downwards gust encounters. The $\alpha^\text{ref.}=5^{\circ}$ cases test the controller's ability to regulate lift about a steady-state attached flow condition while the $\alpha^\text{ref.}=15^{\circ}$ cases test the controller's ability to track lift about a steady-state fully separated flow condition. Results are presented in terms of convective time $t^*=t U_\infty /  c$.

\subsection{Variation in gust ratio $\text{GR}$}
\begin{figure}[h]
     \centering
         \includegraphics[width=0.5\linewidth]{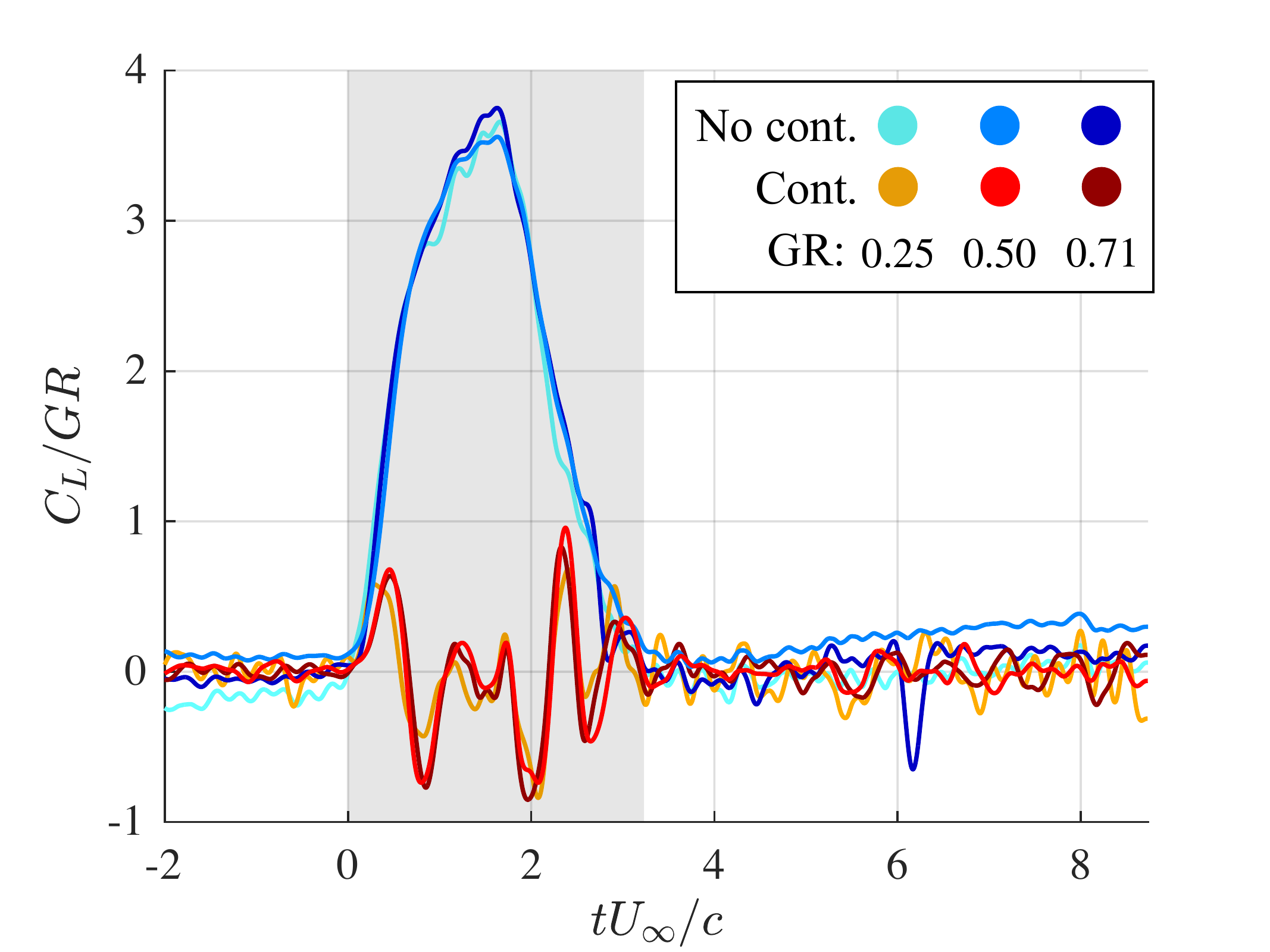}
         \caption{Coefficient of lift normalized by the gust ratio $C_L/\text{GR}$ of the no-control and closed-loop-control cases for various gust ratios at a reference angle of attack $\alpha^\text{ref.}=0$}  
          \label{fig:Lift_AoA_0}   
\end{figure}

Figure~\ref{fig:Lift_AoA_0} compares the coefficients of lift between the no-control and closed-loop-control experiments for gust ratios $\text{GR}=0.25$, $0.50$, and $0.71$ at $\alpha^\text{ref.}=0^{\circ}$. All the coefficients of lift in the figure are normalized by the gust ratio $\text{GR}$. The gray region in the plots highlights the time interval during which the wing was at least partially within the gust. Figure~\ref{fig:Lift_AoA_0} demonstrates a collapse of the time-resolved force trends for the controlled and uncontrolled wings which indicates that both force trends scale linearly with $\text{GR}$ up to the highest gust ratio tested. Without control, the wing experiences a lift increase when it enters the gust at $t U_\infty /  c = 0.00$. Lift continues to increase until $t U_\infty /  c = 1.80$, and then begins to decrease. Lift recovers to steady state $C_L=0$ once the wing fully exits the gust. As the gust ratio increases, so does the peak lift the wing reaches.

Under closed-loop control, the wing's lift initially increases as the wing enters the gust. The wing's lift continues to increase until $t U_\infty /  c = 0.36$. Following this local peak in lift, lift decreases until it reaches a local minimum at $t U_\infty /  c = 0.80$. This initial region of the encounter marks the wing's entry period into the gust. At $t U_\infty /  c = 1.00$, the wing is fully immersed in the gust. Between $t U_\infty /  c = 1.00$ and $t U_\infty /  c = 1.80$, the wing experiences minor oscillations in lift about $C_L=0$. At $t U_\infty /  c = 1.80$, the wing's lift begins to decrease to another local minimum as the wing starts to exit the gust. This local minimum is followed by a local peak at $t U_\infty /  c = 2.4$ and a return to $C_L=0$ outside the gust. All wing-gust encounters exhibit a similar oscillation trend regardless of gust ratio. While the controller is successful at mitigating the lift transient, it performs worse than the simulated feedback system within the Wagner-K\"ussner framework (figure~\ref{fig:Kussner}). Pitching moments transients are also important for vehicle stability. While the pitching moments are outside of the scope of this work, the reader is referred to the authors' previous work which presents and discusses the pitching moment transients in the case of open-loop pitch maneuvers in a transverse gust~\cite{sedky2022physics}.

\begin{figure}[h]
\centering
\subfloat[0.32\textwidth][$\text{GR}=0.25$]{\includegraphics[width=0.32\textwidth]{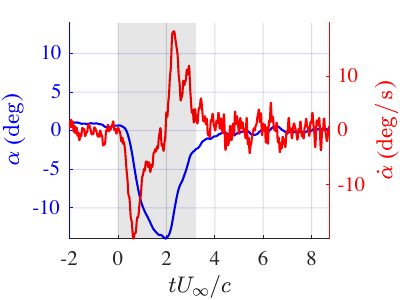}}\hfill
\subfloat[0.32\textwidth][$\text{GR}=0.50$]{\includegraphics[width=0.32\textwidth]{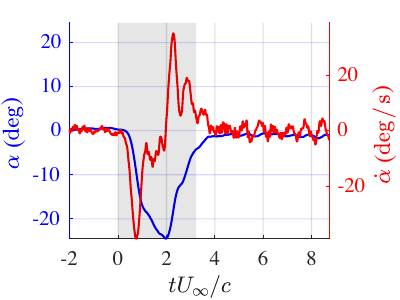}}\hfill
\subfloat[0.32\textwidth][$\text{GR}=0.71$]{\includegraphics[width=0.32\textwidth]{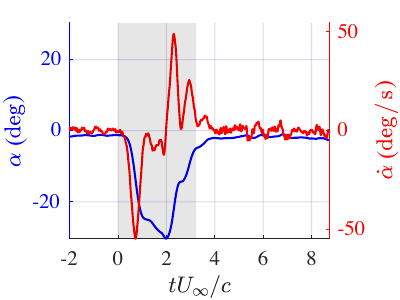}}\hfill
 \caption{Angle of attack $\alpha$ ({\color{blue} blue})  and pitch rate $\dot{\alpha}$ ({\color{red} red}) for closed-loop control for various gust ratios at a reference angle of attack $\alpha^\text{ref.}=0$}
      \label{fig:Kinematics_AoA_0}
\end{figure}

Figure~\ref{fig:Kinematics_AoA_0} presents the corresponding angle of attack $\alpha$ and pitch rate $\dot{\alpha}$ for the closed-loop control experiments. As the wing enters the gust, it pitches down to a minimum $\alpha$ at $t U_\infty /  c = 2.1$, and then pitches back up to $\alpha=0$ as the wing exits the gust. Note that the wing in all cases does not begin to pitch immediately. Rather, it begins to pitch at $t U_\infty /  c = 0.25$. Since the closed-loop controller based on lift measurements needs a lift variation to react to, the lift increases slightly before the wing reacts to it. 

An understanding of the oscillatory trend in the controlled cases can be gained by inspecting the wing's kinematics shown in figure~\ref{fig:Kinematics_AoA_0}. During the gust entry region ($t U_\infty /  c = 0.00$ to $t U_\infty /  c = 0.36$), there is a delay in the wing's pitch-down motion, which is due to the use of feedback  control and the application of a low-pass moving-average filter to the sensor measurements. This delay coincides with the higher lift peak at $t U_\infty /  c = 0.36$. Subsequently, the wing over corrects for the delayed pitch down motion by pitching down aggressively, leading to the local minimum at $t U_\infty /  c = 0.80$. The high negative pitch rate $\dot{\alpha}$, along with the associated lower angle of attack $\alpha$, lead to elevated negative circulatory and added-mass forces, which cause the local minimum. These force contributions are discussed in greater detail in Section~\ref{Sec:ForceDecomposition}. This delay also decreases the controller's performance relative to the simulated controller in the Wagner-K\"ussner framework. Between $t U_\infty /  c = 1.00$ and $t U_\infty /  c = 1.80$, the wing becomes fully immersed in the gust. Since the flow conditions in this region are relatively constant compared to gust entry and exit regions, the wing's coefficient of lift exhibits minimal excursions from the reference coefficient of lift $C_L=0$. As the wing starts to exit the gust at $tU/c = 2.00$, it starts to pitch back to its reference angle of attack. The wing exhibits a delayed response in its pitch up motion. The delayed response in pitching up as the wing exits the gust leads to negative lift. Subsequently, the wing over corrects by aggressively pitching up, leading to elevated positive circulatory and added mass forces, and a corresponding local peak in lift at $t U_\infty /  c = 2.4$.

\begin{figure}[h]
     \centering
         \includegraphics[width=\linewidth]{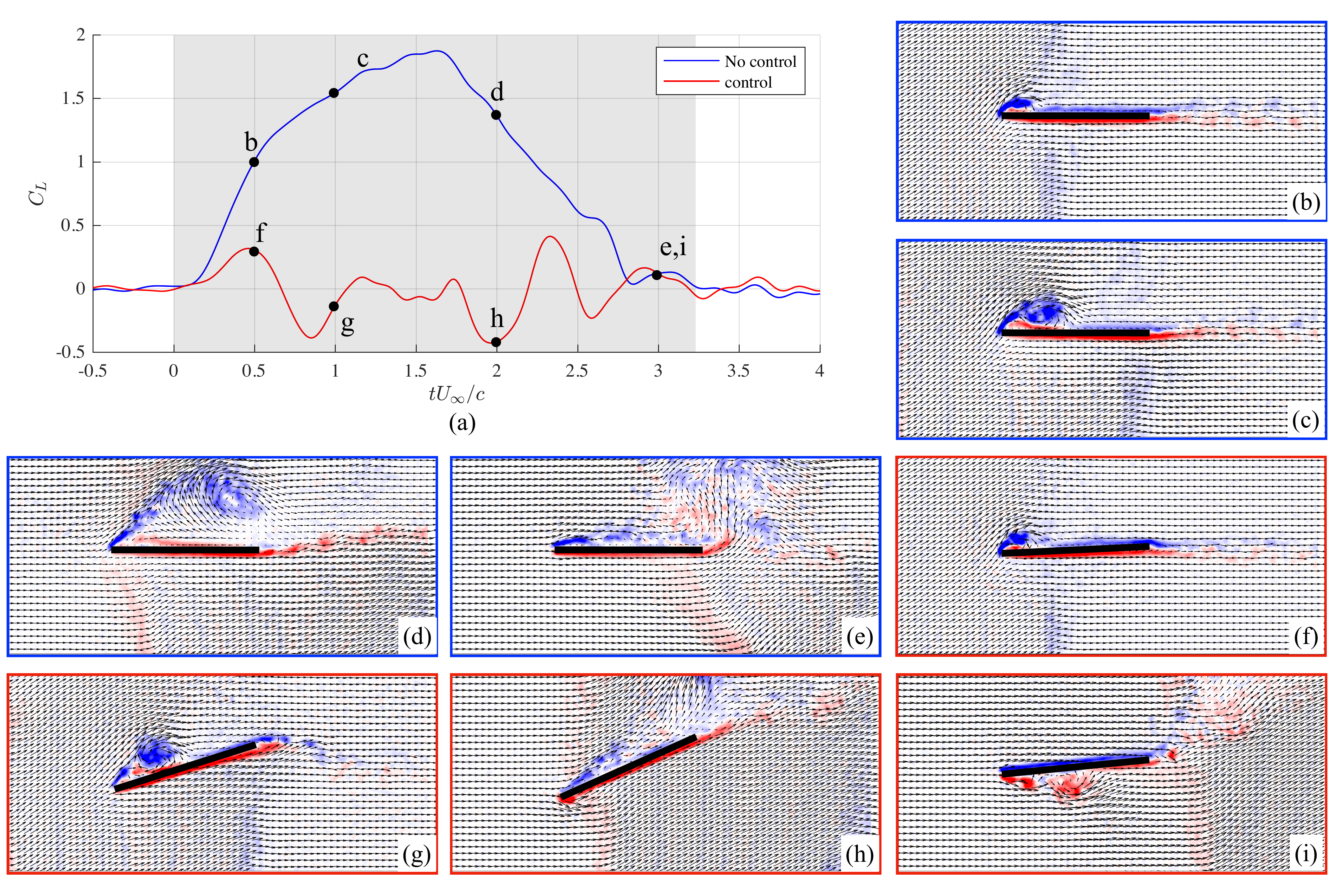}
 \caption{(a) Lift coefficients for reference angle of attack $\alpha^\text{ref.}=0^{\circ}$ and gust ratio $\text{GR}=0.50$. The velocity and vorticity fields for the no control and control cases at $t^*=0.50$ (b, f),  $t^*=1.00$ (c, g), $t^*=2.00$ (d, h), and $t^*=3.00$ (e, i), respectively.}
          \label{fig:Flowfields_GR_0.50}  
\end{figure}

Figure~\ref{fig:Flowfields_GR_0.50} shows the flowfields for wing-gust encounter experiments with and without closed-loop control for gust ratio $\text{GR}=0.50$. As the wing without control enters the gust, the upwash creates a strong leading-edge shear layer and a subsequent LEV (e.g., see figure~\ref{fig:Flowfields_GR_0.50}b at $t U_\infty /c=0.50$). As the vortex grows, it induces the formation of secondary vorticity under it (e.g., see figure~\ref{fig:Flowfields_GR_0.50}c at $t U_\infty /c=1.00$). The shear layer continues to feed vorticity to the LEV, and the vortex grows until it separates (e.g., see figure~\ref{fig:Flowfields_GR_0.50}d at $t U_\infty /c=2.00$). The lift decrease coincides with the detachment of the LEV. The trailing-edge wake of the non-maneuvering wing experiences an upwash from the gust, causing it to bend slightly upwards. 

In contrast, as the wing with closed-loop control enters the gust, it pitches down. An LEV develops, like the encounter without pitching (e.g., see figure~\ref{fig:Flowfields_GR_0.50}f at $t U_\infty /c=0.50$), but the LEV of the maneuvering wing detaches sooner than the LEV of the non-maneuvering wing (e.g., see figure~\ref{fig:Flowfields_GR_0.50}e at $t U_\infty /c=1.00$). As the wing exits the gust, it pitches up and an opposite-signed LEV forms on the high-pressure side of the wing. The high-pressure-side vortex sheds, and at $t U_\infty /c=3.00$ a new one forms, as shown in figure~\ref{fig:Flowfields_GR_0.50}i. 

During the pitch motion, added mass plays a leading role in the lift oscillations observed in the controlled case. While the LEV is known to augment lift~\cite{ford2013lift}, the lift experienced by the controlled wing dips below zero during the  wing's entry into the gust even during the presence of a strong LEV on the wing's suction side (e.g., see figure~\ref{fig:Flowfields_GR_0.50}g). The dip in lift can be attributed to the presence of strong negative added mass force contribution due to the rapid pitch-down rate during this period. However, section~\ref{Sec:ForceDecomposition} also shows how the circulatory contribution to lift can contribute to the lift dip in the presence of a strong LEV. 

\subsection{Variation in the reference angle of attack $\alpha^\text{ref.}$}
\begin{figure}[h]
\centering
\subfloat[0.32\textwidth][$\alpha^\text{ref.}=5^{\circ}$]{\includegraphics[width=0.32\textwidth]{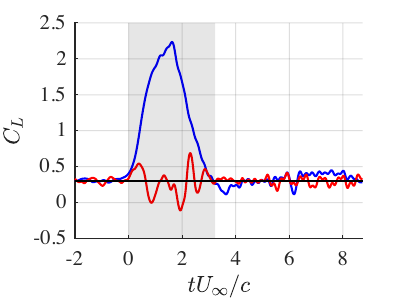}}\hfill
\subfloat[0.32\textwidth][$\alpha^\text{ref.}=10^{\circ}$]{\includegraphics[width=0.32\textwidth]{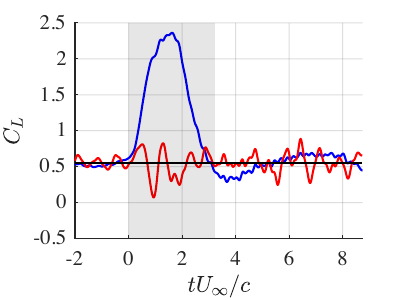}}\hfill
\subfloat[0.32\textwidth][$\alpha^\text{ref.}=15^{\circ}$]{\includegraphics[width=0.32\textwidth]{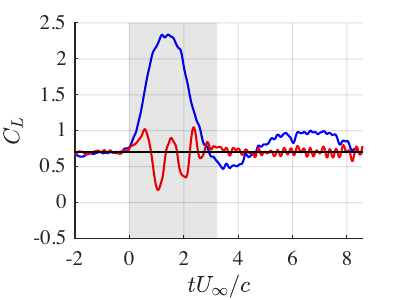}}\hfill
 \caption{A comparison of the coefficient of lift $C_L$ time histories between no control ({\color{blue} blue}) and closed-loop control ({\color{red} red}) for $\text{GR}=0.50$ and various reference angles of attack.}
      \label{fig:Lift_HighAoA}
\end{figure}

This section presents experimental results for closed-loop lift tracking at non-zero reference angles of attack $\alpha^\text{ref.}=5^{\circ}$, $10^{\circ}$, and $15^{\circ}$ and gust ratio $\text{GR}=0.50$. The goal of the controller in these cases is to track a reference lift value that corresponds to the angle of attack $\alpha^\text{ref.}$ in steady state, i.e., prior to gust entry. Figure~\ref{fig:Lift_HighAoA} plots the lift coefficients of the wing with and without closed-loop control for various reference angles of attack at $\text{GR}=0.50$. The horizontal black line is the reference lift signal. There are two main regions in the force histories: the main gust and the gust recovery periods. The main gust period is the period during which the wing is at least partially immersed in the gust, and it spans $t U_\infty /c =0.00$ to $t U_\infty /c =3.23$. The gust recovery period is the period after the wing fully exits the gust but still experiences significant lift transients, spanning $t U_\infty /c =3.23$ onward. The lift transients in the gust recovery region have been shown to be large at high angles of attack~\cite{biler2018experimental,perrotta2017unsteady,andreu2020effect}. The lift transients in the gust recovery region consist of an initial dip and a secondary peak in lift, as shown in figure~\ref{fig:Lift_HighAoA}. Perrotta and Jones attributed the force reduction as the wing exits the gust to the formation of a large trailing edge vortex, which is supported by the flowfield measurements of this section

The lift of the no-control case starts rising from its steady state value at $t U_\infty /c =0.00$. Lift continues to increase in the main gust region until $t U_\infty /  c = 1.80$, and then begins to decrease until the wing fully exits the gust. These trends are similar to the trends seen in figure~\ref{fig:Lift_AoA_0}. The main difference between the wing-gust encounters at $\alpha^\text{ref.}=0$ and $\alpha^\text{ref.}>0$ is the presence of a transient in the gust recovery region. The wing experiences a dip in lift soon after exiting the gust ($t U_\infty /  c = 3.50$). The dip in force is followed by a secondary peak ($t U_\infty /  c = 6.00$). The flowfields presented in this section will explain the lift response in the gust recovery region. 

The lift of the wing with closed-loop control remains close to the reference lift value prior to $t U_\infty /c =0.00$ for all tested reference angles of attack, demonstrating that the controller can successfully track positive steady-state lift at pre- and post-stall angles of attack. Oscillations within the gust region, like those at $\alpha^\text{ref.}=0$, can be observed. The lift of the wing with closed-loop control remains close to the reference lift in the gust recovery region. Thus, the controller is shown to successfully mitigate the gust's main transient as well as the secondary transient in the gust recovery region.

\begin{figure}[h]
\centering
\subfloat[0.32\textwidth][$\alpha^\text{ref.}=5^{\circ}$]{\includegraphics[width=0.32\textwidth]{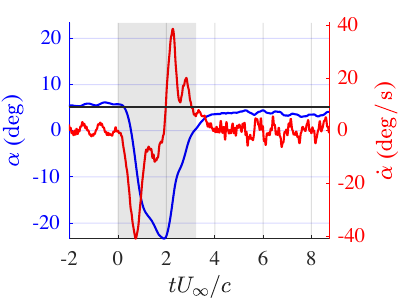}}\hfill
\subfloat[0.32\textwidth][$\alpha^\text{ref.}=10^{\circ}$]{\includegraphics[width=0.32\textwidth]{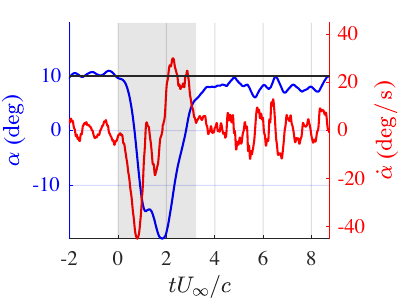}}\hfill
\subfloat[0.32\textwidth][$\alpha^\text{ref.}=15^{\circ}$]{\includegraphics[width=0.32\textwidth]{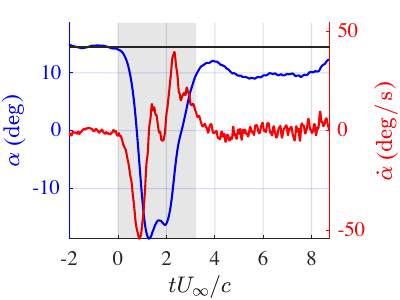}\label{Kinematics_HighAoA_c}}\hfill
 \caption{Angle of attack $\alpha$ ({\color{blue} blue})  and pitch rate $\dot{\alpha}$ ({\color{red} red}) for closed-loop control at $\text{GR}=0.50$ and various reference angles of attack.}
       \label{fig:Kinematics_HighAoA}
\end{figure}

Figure~\ref{fig:Kinematics_HighAoA} presents the corresponding angle of attack $\alpha$ and pitch rate $\dot{\alpha}$ for the closed-loop-control experiments. The horizontal black line is the reference angle of attack, i.e., the angle of attack that corresponds to the non-zero reference lift coefficient. Like the $\alpha^\text{ref.}=0$ cases, the wing pitches down from its steady-state value as it enters the gust then pitches back up as it exits the gust. An important distinction between the $\alpha^\text{ref.}>0$ and $\alpha^\text{ref.}=0$ is that the angle of attack $\alpha$ remains below the steady-state reference value in the gust recovery region for the $\alpha^\text{ref.}>0$ cases, as shown in figure~\ref{Kinematics_HighAoA_c}. The lower angle of attack reduces lift in the gust recovery region preventing the development of the secondary lift peak observed in the cases without control. 

\begin{figure}[h]
     \centering
         \includegraphics[width=\linewidth]{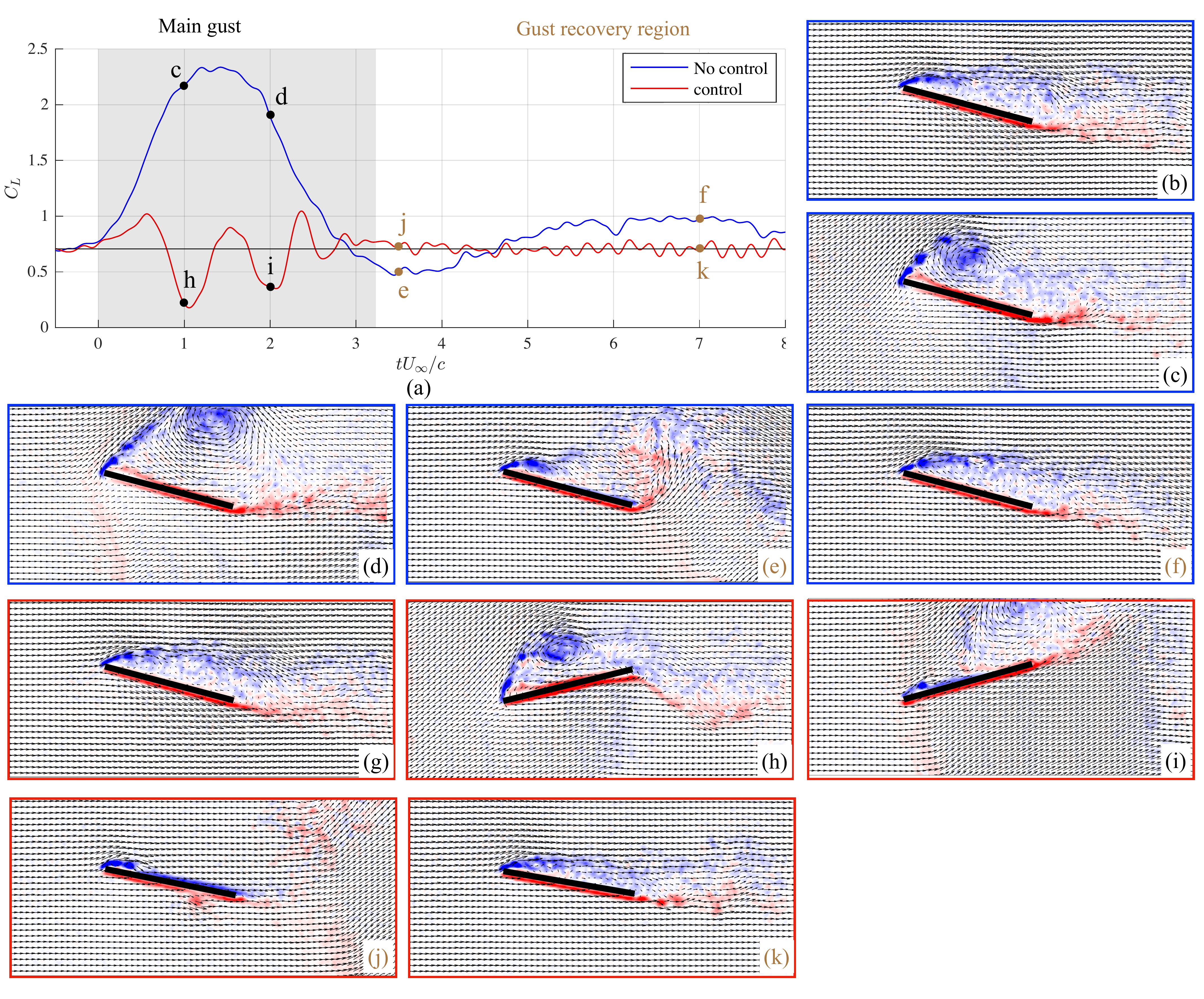}
 \caption{(a) Lift coefficient for $\alpha^\text{ref.}=15^{\circ}$ and $\text{GR}=0.50$. The velocity and vorticity fields for the no control and control cases at $t^*=-2.00$ (b, g),  $t^*=1.00$ (c, h), $t^*=2.00$ (d, i), $t^*=3.50$ (j, e), and $t^*=7.00$ (f, k), respectively.}
          \label{fig:Flowfield_AoA_15}  
\end{figure}

Figure~\ref{fig:Flowfield_AoA_15} shows the flowfields for wing-gust encounter experiments with and without closed-loop control for reference angle of attack $\alpha_\text{ref.}=15^{\circ}$. At $\alpha_\text{ref.}=15^{\circ}$, the flow is initially fully separated. This is demonstrated by the large wake left by the wing at $t U_\infty /  c =-2.00$ prior to gust entry, as shown in figure~\ref{fig:Flowfield_AoA_15}b. Like the $\alpha_\text{ref.}=0$ cases, as the wing without control enters the gust, an LEV forms on its suction side (e.g., see figure~\ref{fig:Flowfield_AoA_15}c at $t U_\infty /  c =0.50$). The formed LEV induces secondary vorticity under it. As the wing starts exiting the gust, the LEV detaches (e.g., see figure~\ref{fig:Flowfield_AoA_15}d at $t U_\infty /  c =2.00$). In the initial period of gust recovery (e.g., see figure~\ref{fig:Flowfield_AoA_15}e at $t U_\infty /c =3.50$), the wing without control exhibits a flow roll-up at the trailing edge, i.e., a trailing-edge vortex, and a nascent leading-edge vortex. While the LEV can be seen to reattach the flow to the wing's suction side, the trailing-edge vortex redirects that flow upwards. As a result, the overall flow is not effectively deflected by the wing's suction side relative to the steady-state freestream case prior to the gust encounter, as shown in figure~\ref{fig:Flowfield_AoA_15}b. The lack of deflection results in a lower lift value compared to steady state. At $t U_\infty /c =7.00$, the uncontrolled wing experiences a secondary peak in lift. This secondary peak corresponds to the formation of a leading-edge vortex that deflects the flow above the suction side of the wing downwards. The wake behind the wing observed at $t U_\infty /c =6.00$, shown in figure~\ref{fig:Flowfield_AoA_15}f, deflects downward more than the wake in the fully stalled steady-state uncontrolled wing case at $t U_\infty /c =-2.00$, shown in figure~\ref{fig:Flowfield_AoA_15}b, which explains the higher lift in the secondary peak region.

As the wing with closed-loop control enters the gust, it pitches down from its steady-state angle of attack. The LEV detaches faster compared to the case without control. As the wing exits the gust, it pitches up, and a small pressure-side LEV forms. As the wing with closed-loop control enters the gust, it pitches down and an LEV forms. The formed LEV sheds earlier than the case without closed-loop control, just as observed in the $\alpha^\text{ref.}=0$ cases. As the wing exits the gust, it pitches up and a pressure-side LEV forms. The pressure-side LEV sheds and convects away at $t U_\infty /  c =3.00$. As discussed earlier, for high reference angles of attack $\alpha^\text{ref.}$ cases, a transient in lift can be observed in the gust recovery region. The closed-loop control strategy successfully mitigates the transient in the recovery region. The wing with closed-loop control exhibits no trailing-edge vortex, and thus, it has a higher lift value where the uncontrolled wing experienced a dip at $t^*=3.50$. In fact, in this flow condition, the nascent LEV attaches the flow, making the lift value higher than the steady-state reference lift. The wing compensates for this fact by maintaining an angle of attack lower than the reference angle of attack, as shown in figure~\ref{Kinematics_HighAoA_c}, and thus the lift of the controlled wing at $t U_\infty /c =3.50$ remains close to the reference lift.

\subsection{Wing encountering downward transverse gust}
\begin{figure}[h]
\centering
\subfloat[0.33\textwidth][$\alpha^\text{ref.}=5^{\circ}$]{\includegraphics[width=0.33\textwidth]{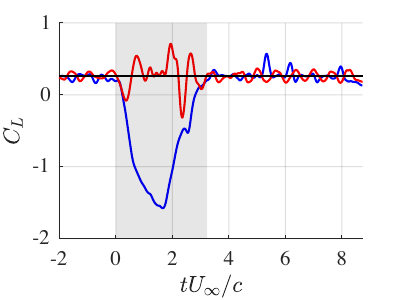}}
\subfloat[0.33\textwidth][$\alpha^\text{ref.}=15^{\circ}$ ]{\includegraphics[width=0.33\textwidth]{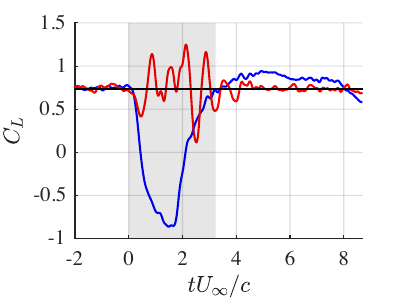}}
 \caption{A comparison of the lift coefficients for no control ({\color{blue} blue}) and closed-loop control ({\color{red} red}) at various reference angles of attack and a downward gust with $\text{GR}=0.50$.}
      \label{fig:Lift_DownGust}
\end{figure}

Figure~\ref{fig:Lift_DownGust} shows the lift coefficient of a wing encountering a downward gust with and without closed-loop control for reference angles of attack $\alpha^\text{ref.}=5^{\circ}$ and $15^{\circ}$, at $\text{GR}=0.50$. The horizontal black line is the reference lift signal. As with the cases of upward gust encounters, the $\alpha^\text{ref.}=15^{\circ}$ exhibits a significant gust recovery region transient. Since the wing is encountering a downward gust, the lift of the no-control case starts decreasing from its steady-state value at $t U_\infty /c =0.00$. Lift continues to decrease in the main gust region until $t U_\infty /  c = 1.80$, and then it begins to increase until the wing fully exits the gust. For the $\alpha^\text{ref.}=5^{\circ}$, the wing's lift returns to steady state once the wing fully exits the gust, while for the $\alpha^\text{ref.}=15^{\circ}$, the wing experiences a significant lift transient in the gust recovery region. While the lift transient of the upward gust was characterized with an initial dip followed by a secondary peak in lift, the downward gust encounter exhibits a peak without a prior dip in lift. It is important to note that a lift dip is observed after $t U_\infty /c =8.00$, but the measurement domain of the towing tank facility did not extend far enough to adequately investigate this region. The flowfields presented in this section will shed light on the reasons behind the unique lift response of the downward-gust recovery region. The lift of the wing with closed-loop control experiences oscillations within the gust region like those presented in the other cases. As the wing enters the gust, it experiences a dip in lift followed by a peak as the wing reacts to the downward gust. Within the center region of the gust, the wing experiences minimal excursions from the reference lift. As the wing exits the gust, it experiences another transient. The lift peaks and then dips as the wing reacts to the change in flow conditions it perceives upon exiting the gust. The $\alpha^\text{ref.}=15^{\circ}$ case performs worse than the $\alpha^\text{ref.}=5^{\circ}$ case, which indicates that the proportional control's ability to regulate the lift may decline at higher angles of attack. The lift of the wing with closed-loop control remains close to the reference lift in the gust recovery region, similar to the case of upward gust encounters, indicating that the controller is good at mitigating the gust recovery region transients independent of the gust's direction.

\begin{figure}[h]
\centering
\subfloat[0.5\textwidth][$\alpha^\text{ref.}=5^{\circ}$]{\includegraphics[width=0.32\textwidth]{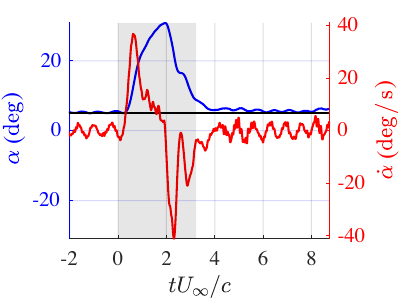}}
\subfloat[0.5\textwidth][$\alpha^\text{ref.}=15^{\circ}$]{\includegraphics[width=0.32\textwidth]{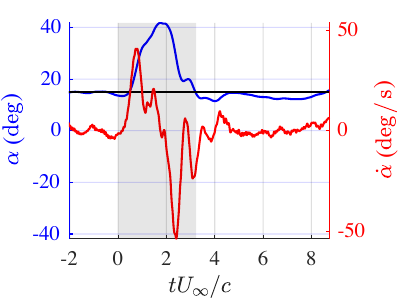}}
 \caption{Angle of attack $\alpha$ ({\color{blue} blue})  and pitch rate $\dot{\alpha}$ ({\color{red} red}) for closed-loop control at various reference angles of attack and a downward gust with $\text{GR}=0.50$.}
       \label{fig:Kinematics_DownGust}
\end{figure}

Figure~\ref{fig:Kinematics_DownGust} presents the corresponding angle of attack $\alpha$ and pitch rate $\dot{\alpha}$ for closed-loop-control experiments. In contrast to the previous cases presented, the wing in the downward gust pitches up and then pitches down. Since the downward gust leads to a reduction in lift, the wing pitches up to compensate for this reduction. In the case of $\alpha^\text{ref.}=5^{\circ}$, the wing returns to its steady-state angle of attack once it exits the gust. In addition, the pitch rate $\dot{\alpha}$ oscillates about the steady state outside of the gust region as a reaction to minor oscillations in lift outside of the gust. In the case of $\alpha^\text{ref.}=15^{\circ}$, the wing maintains an angle of attack lower than the reference angle when the wing exits the gust. The lower reference angle of attack compensates for the secondary lift peak in this case.

\begin{figure}[h]
     \centering
         \includegraphics[width=\linewidth]{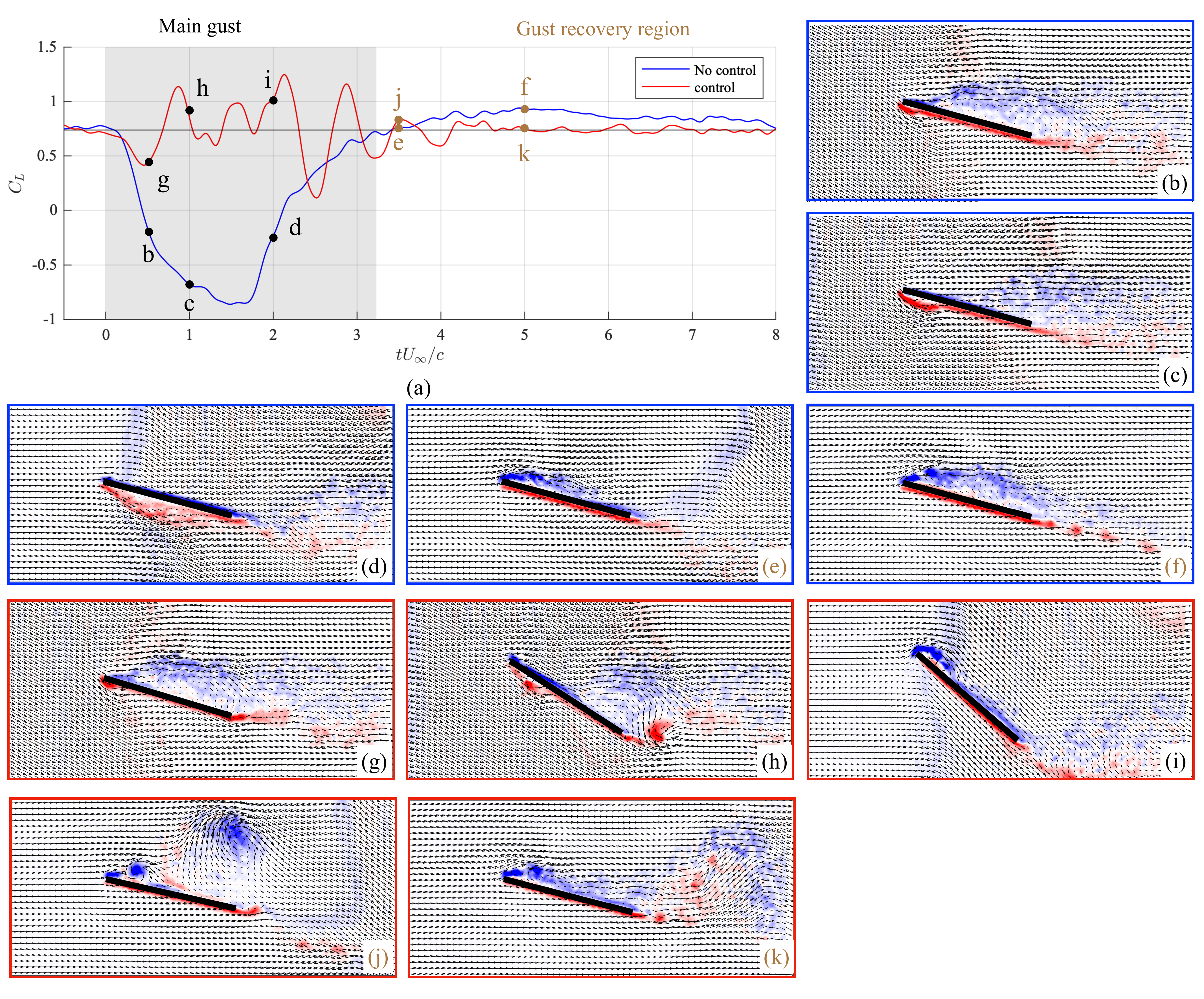}
          \caption{(a) Lift coefficient for $\alpha^\text{ref.}=15^{\circ}$ and a downward gust with $\text{GR}=0.50$. The velocity and vorticity fields for the no control and control cases at $t^*=-2.00$ (b, g),  $t^*=1.00$ (c, h), $t^*=2.00$ (d, i), $t^*=3.50$ (e, j), and $t^*=7.00$ (f, k), respectively.}
          \label{fig:Flowfield_AoA_15_DownwardGust}  
\end{figure}

Figure~\ref{fig:Flowfield_AoA_15_DownwardGust} shows the flowfields for wing-gust encounter experiments with and without closed-loop control for reference angle of attack $\alpha_\text{ref.}=15^{\circ}$ and a $\text{GR}=0.50$ downward gust. As in the case with the upward gust, at $\alpha_\text{ref.}=15^{\circ}$, the flow over the wing is fully separated prior to entering the gust. In contrast to the upward gust, as the uncontrolled wing enters the gust, the downward velocity of the gust reattaches the flow on the wing's suction side and creates an LEV on the high-pressure side, (e.g., see figure~\ref{fig:Flowfield_AoA_15_DownwardGust}b at $t U_\infty /  c =0.50$). The vortex grows until $t U_\infty /  c =2.00$ in figure~\ref{fig:Flowfield_AoA_15_DownwardGust}d. The duration of the vortex growth corresponds to the duration of lift reduction. In figure~\ref{fig:Flowfield_AoA_15_DownwardGust}e, as the wing exits the gust, the high-pressure-side vortex detaches, and a new suction-side vortex begins to form. In contrast to the $\alpha^\text{ref.}=15^{\circ}$ wing encountering an upward gust, no flow roll-up is observed at the trailing edge. The absence of a trailing-edge vortex explains the absence of a lift dip during the gust recovery region. In the case of the downward gust encounter, the nascent LEV successfully reattaches the flow to the suction side of the wing, which increases lift and results in a secondary lift peak at $t^*=5.00$. Following the secondary peak, the lift drops as the flow slowly reverts to its steady fully separated state.

As the controller wing with closed-loop control enters the gust, it pitches up to compensate for the reduction in lift from the downward gust. The delay in pitch leads to a lift reduction, and the following reaction to the lift peak results in a lift peak during the gust entrance region. As the wing pitches up, the wing's chord begins to align with the flow and the primary high-pressure-side vortex detaches. At $t U_\infty /  c =2.00$ in figure~\ref{fig:Flowfield_AoA_15_DownwardGust}i, the wing exits the gust at a very high angle of attack, and thus, a suction-side LEV develops. The formation of this LEV coincides with a lift peak in the forces. This vortex detaches, and a new vortex begins to form at $t U_\infty /  c =3.00$, as shown in figure~\ref{fig:Flowfield_AoA_15_DownwardGust}j. The flow topology for the controlled wing encountering the downward gust at $t U_\infty /c =3.50$, shown in figure~\ref{fig:Flowfield_AoA_15_DownwardGust}j, is quite different from the uncontrolled one. The strong suction-side LEV that formed during the gust exit sheds away and another forms. The flow is not fully separated as in the steady-state case, and thus the wing pitches to an angle of attack lower than the reference angle to maintain low lift. The controlled wing maintains an angle of attack lower than the reference angle of attack for most of the gust recovery region to maintain the correct amount of lift until the wing is fully stalled again.

\subsection{Performance quantification}
Let the lift coefficients for the uncontrolled and controlled wings be $C_L^\text{nc}$ and $C_L^\text{c}$, respectively. A mitigation performance measure $\eta$ can be defined to be
\begin{equation}
\eta=\frac{||C_L^\text{nc} - C_L^\text{ref.}||_2 - ||C_L^\text{c} - C_L^\text{ref.}||_2}{ ||C_L^\text{nc} - C_L^\text{ref.}||_2} \times 100,
\end{equation}
where $C_L^\text{ref.}$ is the reference coefficient of lift and $||C_L - C_L^\text{ref.}||_2$ is the Euclidean norm of the vector of discrete-time errors in the lift coefficient from the reference. The mitigation measure $\eta$ tells the percentage reduction in the cumulative deviation of lift from the reference value over the entire gust encounter. This measure is similar to one used by Herrmann et al.~\cite{Herrmann2022Gust} to quantify their controller's performance. 

\begin{table}[h]
\centering
\small
\caption{Summary of the controller's performance for the tested parameter space}
\label{tab:Performance}
\begin{tabular}{ccccc}
\toprule
\toprule
Direction & $\alpha^{\text{ref}}$ & $\text{GR}$ & $\eta$ \% \\ \midrule
Upward    & $0^{\circ}$                              & 0.25        & 82.60         \\
Upward    & $0^{\circ}$                              & 0.50        & 81.96         \\
Upward    & $0^{\circ}$                              & 0.71        & 81.92         \\
Upward    & $5^{\circ}$                              & 0.50        & 83.33        \\
Upward    & $10^{\circ}$                             & 0.50        & 79.01         \\
Upward    & $15^{\circ}$                             & 0.50        & 78.63         \\
Downward  & $5^{\circ}$                              & 0.50        & 80.14         \\
Downward  & $15^{\circ}$                             & 0.50        & 72.33         \\ 
\bottomrule
\bottomrule
\end{tabular}%
\end{table}

Table~\ref{tab:Performance} summarizes the controller's mitigation performance in the closed-loop control experiments. Despite the simplicity of a proportional feedback-control law, the controlled wing provided an average rejection performance of $\eta=80 \%$. Nevertheless, it remained lower than the Wagner-K\"ussner simulated controller performance, $\eta=92 \%$, presented in figure~\ref{fig:Kussner}. As discussed earlier, one of the main contributors to this discrepancy is the presence of a low-pass filter on the feedback signal that delays the wing's reaction to the gust. However, there may be other contributors to this discrepancy such as the nonlinear flow physics caused by the large vortex structures and gust shear layer deformation. The performance of the controller was consistent across gust scenarios that differed based on gust strength, reference angle of attack, and gust direction. The main limitation on the controller's performance is the application of a low-pass filter on the lift measurement signal before feeding it to the controller. The low-pass filter limited the noise impact on the performance but resulted in a delayed reaction to the gust by the wing. The application of the filter was necessary due to the high amplitude of the noise content in the sensor measurements.

\subsection{Force decomposition}
\label{Sec:ForceDecomposition}
A pitching wing encountering a gust experiences lift from three dominant sources. The first contribution to lift is the added-mass force due to unsteady pitching. Added-mass force is classically described as the force needed to accelerate the fluid surrounding the body when the body accelerates~\citep{darwin1953note}. The added-mass coefficient of lift $C_L^\text{am}$ of a wing pitching about its midchord based on potential flow theory can be expressed as~\citep{limacher2018generalized,limacher_2021}

\begin{equation}
C_L^\text{am}=\frac{\pi c}{2U_\infty}\dot{\alpha}\left(\cos^2\alpha - \sin^2\alpha \right),
\label{Eq:14}
\end{equation} 
\noindent where the cosine and sine terms account for the large angle of attack displacements that may be undertaken by the wing.

The second contribution to lift is the non-circulatory gust contribution. Like added mass, this force is related to the non-circulatory effect of fluid acceleration around the wing but due to the time rate of change of the velocity field imposed by the gust on the wing. An estimate of the non-circulatory gust contribution based on the gust velocity can be computed through convolution with the non-circulatory force contribution indicial function, similar to K\"ussner's indicial function

\begin{equation}
C_L^{\text{g}}=\frac{4}{U_\infty}\int_0^{s}\frac{\text{d}v_g}{\text{d}\sigma}\zeta(s-\sigma) \text{d}\sigma\cos\alpha, 
\label{Eq:16}
\end{equation} 

\noindent where the cosine term accounts for the reduction of the wing's area projection perpendicular to the gust flow as the wing pitches and the indicial function based on the expression presented by~\citet{andreu2020effect} is 

\begin{equation}
\zeta(s)=\sqrt{\frac{1}{2} s- \frac{1}{4} s^2}.
\label{Eq:17}
\end{equation}

The final contribution to lift is the circulatory contribution due to the production and convection of circulatory structures in the flow, including the wing's wake, leading-edge vortex, and circulatory vortex sheet along the wing. The total coefficient of lift, $C_L$, can be decomposed into an added-mass force contribution, $C_L^\text{am}$, non-circulatory gust contribution $C_L^{\text{g}}$, and a circulatory contribution $C_L^\text{c}$~\citep{CorkeryThesis,corkery_babinsky_graham_2019}, such that

\begin{equation}
C_L=C_L^\text{am}+C_L^{\text{g}}+C_L^\text{c}.
\label{Eq:Decompose}
\end{equation}

\noindent Based on equation~\eqref{Eq:Decompose}, the circulatory component of lift can be computed by subtracting the added-mass force contribution and the non-circulatory gust contribution from the total lift measurement.More on the force decomposition methodology and applicability to current experiments can be found in reference~\cite{sedky2022physics}.

\begin{figure}[h]
\centering
\subfloat[0.5\textwidth][]{\includegraphics[width=0.5\textwidth]{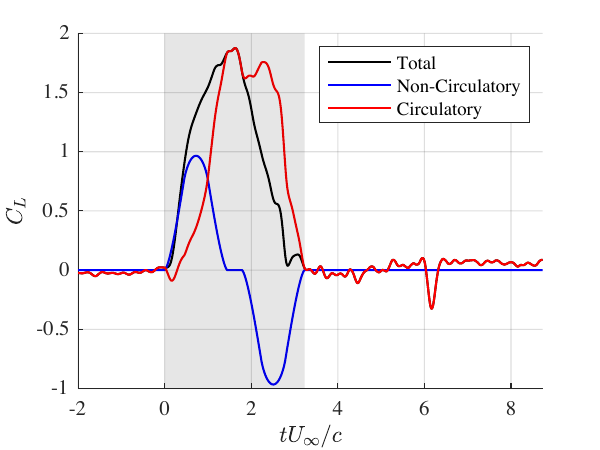}}\hfill
\subfloat[0.5\textwidth][]{\includegraphics[width=0.5\textwidth]{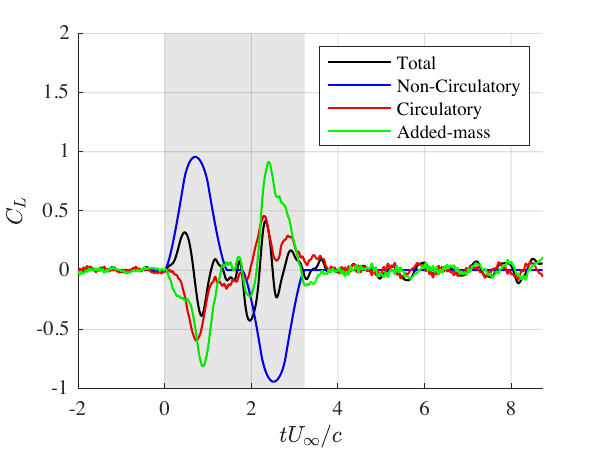}}\hfill
\caption{Lift decomposition of the uncontrolled case (a) and controlled case (b) for the $\alpha^\text{ref.}=0^{\circ}$ and $\text{GR}=0.50$ case.}
        \label{fig:Decomposition}
\end{figure} 

Figure~\ref{fig:Decomposition} shows the lift decomposition for $\alpha^\text{ref.}=0^{\circ}$ and $\text{GR}=0.50$. There is no added-mass force contribution in the no-control case because the wing does not pitch. During entry into the gust, the controlled wing pitches down to reduce the relative flow velocity normal to the chord, and the negative pitch rate leads to a negative added-mass force peak. As the wing exits the gust, it pitches up, and the positive pitch rate leads to a positive added-mass force peak.

Both no control and control cases exhibit similar non-circulatory gust contributions to lift. This contribution is largest during the periods of gust entry and exit when the wing experiences the largest variations in time of velocity imposed on it by the gust. As the wing gradually enters the gust, the vertical velocity the gust induces on the wing increases, which leads to an increase in the non-circulatory gust contribution. The maximum value of this contribution corresponds to the region with the highest rate of change of induced vertical velocity. As the wing exits the gust, the vertical velocity the gust induced on the wing decreases, which leads to a negative non-circulatory gust contribution. The added-mass force acts to counterbalance the non-circulatory force contribution in the closed-loop control case. This demonstrates how the added-mass force produced when kinematic actuation is used as a control input plays an integral role in mitigating the gust. 

The circulatory contributions to lift for the no-control and closed-loop-control cases are different. As the no-control wing enters the gust, a leading-edge vortex forms and continues to accumulate circulation. As the accumulation of circulation in the vortex grows, so does the circulatory contribution to lift. Lift growth due to circulation development is not instantaneous, and thus, circulatory lift takes time to grow. This can be observed in the phase lag between the circulatory and non-circulatory force contributions. On the other hand, the controlled wing minimizes the circulatory force transient.

\begin{figure}[h]
\centering
\subfloat[0.5\textwidth][]{\includegraphics[width=0.5\textwidth]{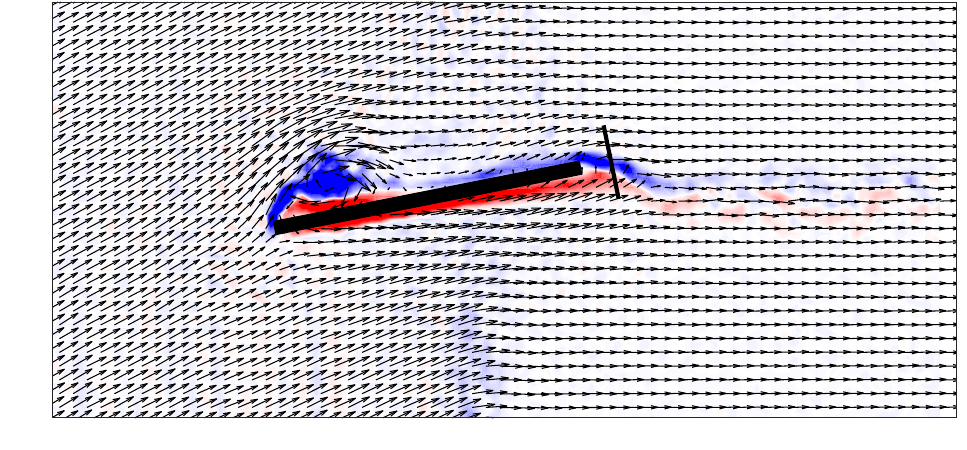}\label{fig:FluxLine_a}}\hfill
\subfloat[0.5\textwidth][]{\includegraphics[width=0.5\textwidth]{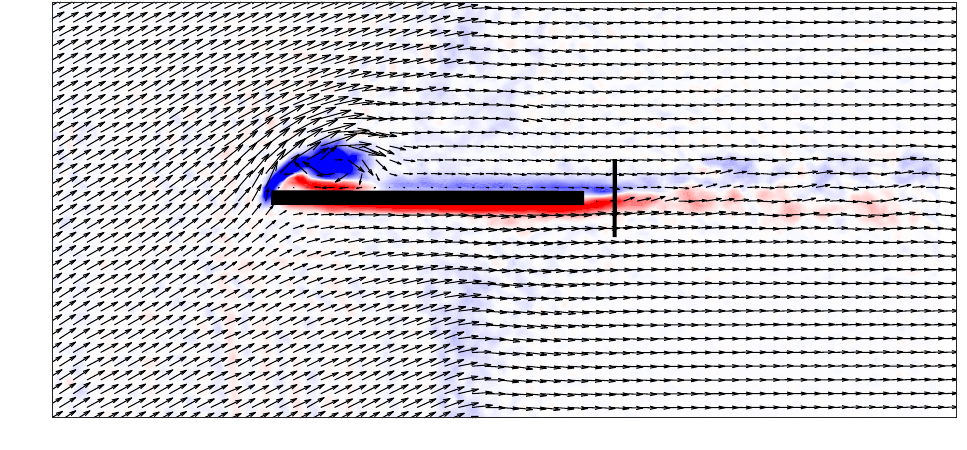}}\hfill
      \caption{Flowfields snapshots and vorticity flux calculation boundary at $t U_\infty /c =0.75$ of the controlled case (a) and uncontrolled case (b) for reference angle of attack $\alpha^\text{ref.}=0^{\circ}$ and gust ratio $\text{GR}=0.50$}
         \label{fig:FluxLine}
\end{figure}

Figure~\ref{fig:Decomposition} shows that the circulatory force contribution of the controlled wing is negative during gust entry even though a strong leading-edge vortex, which should presumably cause a positive circulatory contribution, exists. In fact, a simple inspection of the flowfields of the control and no-control cases presented in figure~\ref{fig:FluxLine} shows a very similar circulation distribution: a strong LEV, some secondary vorticity induced by the LEV, and similar amounts of vorticity in the boundary layers of the wings. As a result, it may be reasonable to assume that the circulatory contribution to lift in both cases would be similar at that instant. However, while the LEV may have similar circulation distribution, the circulation distribution in the boundary layer of the wing may be significantly different. The PIV measurements do not spatially resolve the boundary layer distribution, and thus it may not capture the amount of circulation in the boundary layer. 

\begin{figure}[h]
\centering
         \includegraphics[width=0.7\linewidth]{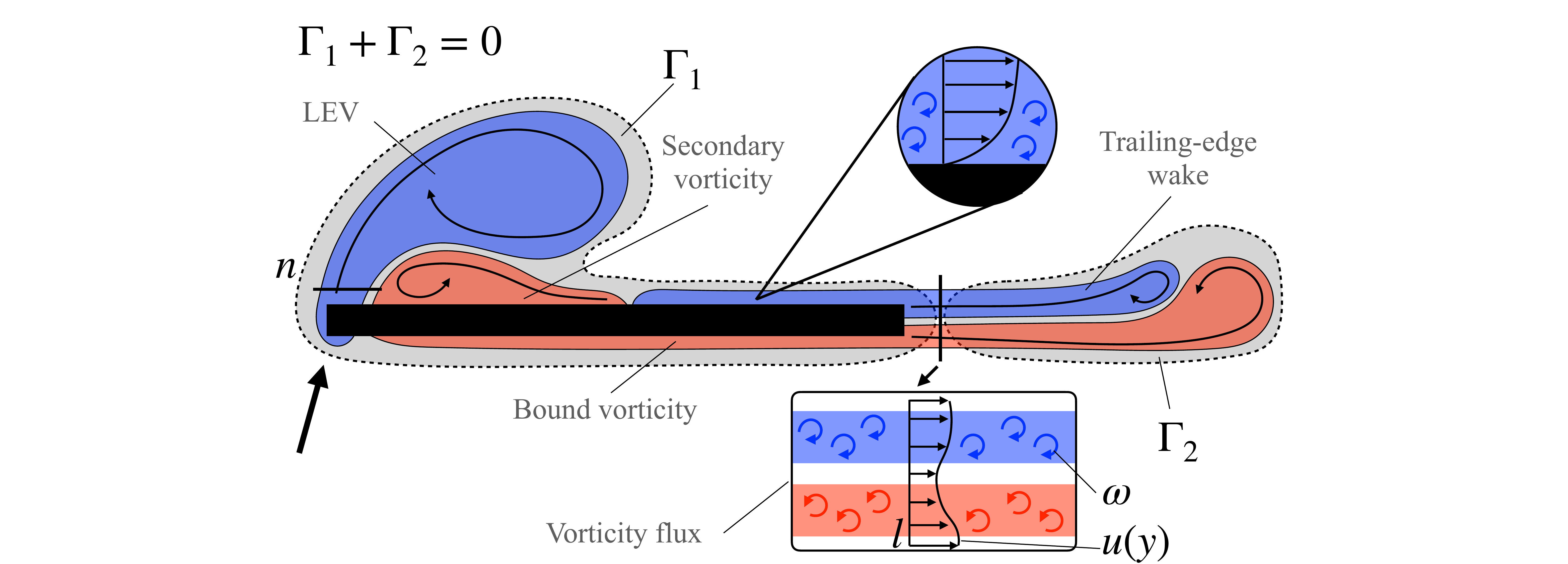}
         \caption{Schematic of circulation distribution in the flowfield of a wing encountering a transverse gust}
         \label{fig:FlowTopologySchematic}
     \end{figure}

To quantify the circulation in the wing's boundary layer, a flux method is used. Figure~\ref{fig:FlowTopologySchematic} shows a general sketch of the vorticity distribution of a wing encountering a gust. The four main regions around the wing where circulation is concentrated are: (1) the LEV containing negative circulation, (2) the secondary vorticity containing positive circulation, (3) the boundary layer, and (4) the trailing-edge wake that contains circulations of different signs. The success of PIV at resolving the vorticity in different regions of the flow is highly dependent on the shear gradient. The boundary layer of the wing contains very high shear gradient due to the wall, and the PIV measurements may not adequately resolve those. In addition, regions around the wing may suffer from laser-light reflections that lead to errors in the PIV correlations close to the wing's boundary. The regions in the flowfield that contain circulation may be divided into two main regions, $\Gamma_1$ which contains the circulations in the LEV, secondary vorticity, and bound vorticity, and $\Gamma_2$ which contains the circulation in the wake shed by the wing.

For an inviscid, incompressible flow, Kelvin's conservation of circulation dictates that circulation is conserved within a closed contour that moves with the fluid~\cite{anderson1984fundamentals}. The same result was shown to hold for viscous flow if the contour chosen does not cross regions with significant shear stresses~\cite{wu1978theory}. Thus, for a flow domain that initially contains zero circulation, i.e., a large flow domain that contains the wing before traveling to encounter the gust, the same flow domain will contain a net zero circulation. While each gust shear layer contains some net circulation, for a symmetric gust, the circulations of each shear layer sum to zero. As such, the remaining regions of circulation, $\Gamma_1$ and $\Gamma_2$, must also sum to zero, such that $\Gamma_1=-\Gamma_2$. Thus, we can quantify the circulation contained in the combined boundary layer, secondary vorticity, and LEV region $\Gamma_1$ by quantifying the circulation contained in the trailing-edge wake. To calculate the time rate of change of circulation in the trailing-edge wake, the vorticity flux from the trailing edge into the wake is integrated along the boundary $l$ shown in figure~\ref{fig:FlowTopologySchematic}.

\begin{equation}
\frac{\text{d} \Gamma_2}{\text{d} t}=\int_l \omega u(y) \text{d}y.
\label{Eq:Flux}
\end{equation}  

\noindent An example of the boundary used is presented in figure~\ref{fig:FluxLine}. Since the wing translates at a constant speed with a zero angle of attack prior to the encounter, the trailing-edge wake is assumed to have a net of zero circulation prior to gust encounter. Thus, the rate of change of circulation can be integrated in time to arrive at the circulation in the trailing-edge wake at a given time. If the wing had a non-zero angle of attack, then the circulation of the starting vortex would have to be taken into account.

\begin{figure}[h]
\centering
\subfloat[0.5\textwidth][]{\includegraphics[width=0.5\textwidth]{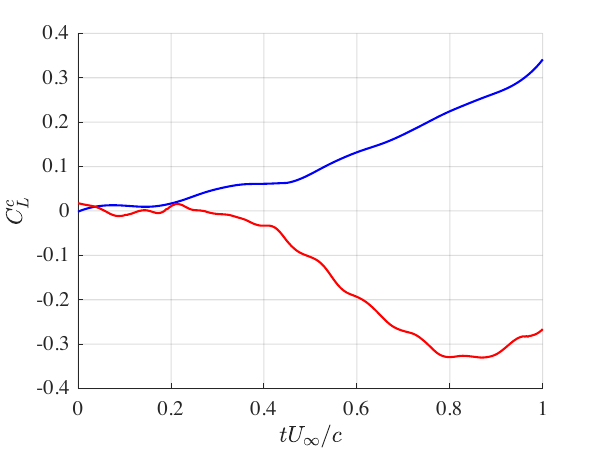}}\hfill
\subfloat[0.5\textwidth][]{\includegraphics[width=0.5\textwidth]{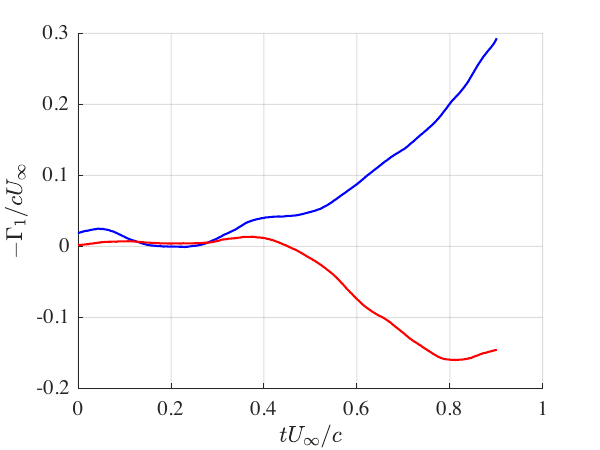}\label{fig:CirculationGamma1}}\hfill
      \caption{A comparison of the circulatory contribution to the coefficient of lift $C_L^c$ (a) as well as the circulation in the $\Gamma_1$ region (b) for no control ({\color{blue} blue}) and closed-loop control ({\color{red} red}) at a reference angle of attack $\alpha^\text{ref.}=0$ and gust ratio $\text{GR}=0.50$.}
      \label{fig:CircLift_Circulation}
\end{figure}

Figure~\ref{fig:CircLift_Circulation} presents the circulatory contribution to the coefficient of lift $C_L^c$ as well as the combined circulation in the LEV, secondary vorticity, and wing's boundary layer $\Gamma_1$ for no control and closed-loop control. The time duration of figure~\ref{fig:CircLift_Circulation} is limited to the entry of the wing into the gust (i.e., for $t U_\infty /c = 0$ to 1), because the LEV sheds in the controlled case at approximately $t U_\infty /c = 0.9$.  The calculation of circulation in the region surrounding the LEV (i.e., $\Gamma_1$) was terminated at $t U_\infty /c = 0.9$ for the uncontrolled case as well. For the case without control, as the wing enters the gust, the circulatory contribution to the lift force increases. Negative circulation here corresponds to clockwise circulation which has a positive contribution to lift. Thus, the negative of the circulation in the $\Gamma_1$ region is plotted in figure~\ref{fig:CircLift_Circulation}. Similar to the circulatory contribution to lift, the negative of the circulation in the $\Gamma_1$ region grows as the wing enters the gust. This is an indication that for this case, the circulation is dominated by the LEV. In this case, it is straightforward to see the correlation between the circulation growth and the growth of the circulatory contribution to lift. In contrast, for the closed-loop control case, both the circulatory contribution to lift as well as the negative of the circulation in the $\Gamma_1$ region decrease as the wing enters the gust. In other words, even though figure~\ref{fig:FluxLine_a} shows massive LEV with negative circulation, the net circulation in the $\Gamma_1$ region which contains the LEV, secondary vorticity, and the boundary layer is positive. This indicates that most of the circulation in this case is not in the LEV, rather it is within the boundary layer and secondary vorticity. 

The flux of negative vorticity through the flux boundary $n$ shown in figure~\ref{fig:FlowTopologySchematic} was used to calculate the growth of the LEV's circulation $\Gamma_\text{LEV}$. Based on Kelvin's conservation of circulation in the flow, the circulation in the boundary layer and secondary vorticity is

\begin{equation}
\Gamma_{\text{SV+BL}}=-\Gamma_\text{LEV}-\Gamma_2.
\end{equation}

\begin{figure}[h]
\centering
         \includegraphics[width=\linewidth]{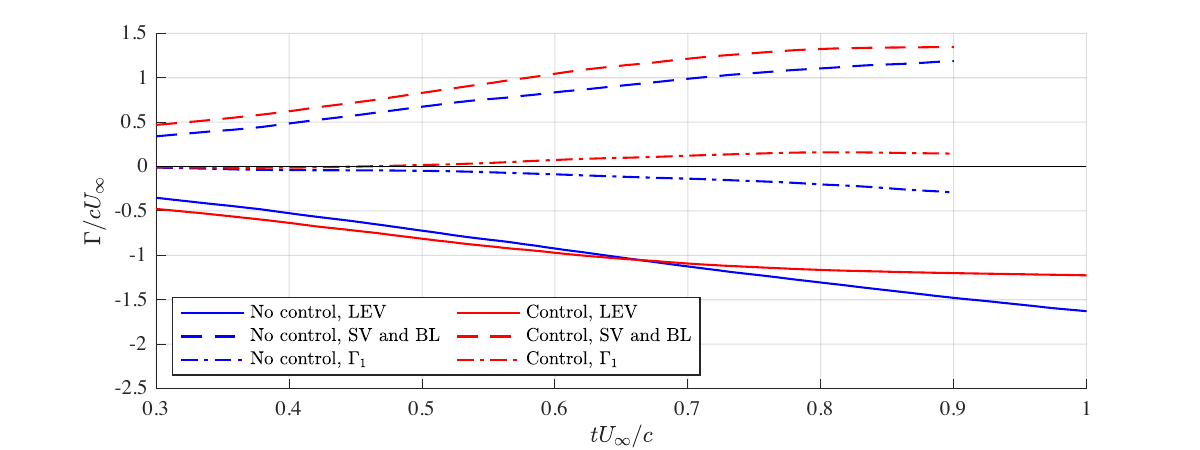}
         \caption{Comparison between the circulation contained in the LEV and the circulation contained in the secondary vorticity and the boundary layer of the wing}
         \label{fig:LEV_Bound_Circulation}
     \end{figure}

\noindent Figure~\ref{fig:LEV_Bound_Circulation} presents a comparison of the circulation within the LEV, the circulation of the secondary vorticity (SV) and the boundary layer (BL), and the $\Gamma_1$ region. The blue trends are for the uncontrolled case and the red trends are for the control case. Note that the sign of the circulations in this figure is not reversed as in figure~\ref{fig:CirculationGamma1}. Similar to figure~\ref{fig:CircLift_Circulation}, some circulation trend lines stop at approximately $t U_\infty /c = 0.9$ due the termination of calculations after LEV shedding in the controlled case. Without control, the LEV's circulation is seen to grow approximately linearly. When control is applied, the LEV of the wing starts at a higher strength. While both wings experience the same gust, upon the controlled wing entry into the gust, it pitches down, moving the leading edge of the wing. The motion of the leading edge adds to the shear layer's velocity, creating a stronger LEV. As the controlled wing lines itself up with the flow, the velocity component of the flow normal to the wing's chord at the leading edge decreases, slowing down the growth of the vortex. Thus, the circulation of the LEV for the two cases start to converge, until the uncontrolled wing's circulation surpasses the controlled wing at $t U_\infty /c=0.67$. Later, the growth of the LEV of the controlled wing plateaus and the LEV detaches. 

While the LEV's circulation of the controlled wing starts off higher but becomes lower than the uncontrolled wing throughout the gust entry period, the circulation within the boundary layer and secondary vorticity of the controlled wing remains higher throughout the entry period. The higher amount of circulation in the SV and BL offset the LEV in the case of the controlled wing which leads to positive circulation in the combined LEV, SV, and BL region ($\Gamma_1$). This results in a negative circulatory contribution to lift as shown earlier.  

The circulatory contribution to lift is governed, not by the circulation of the LEV alone, but by the circulation in the combined LEV, SV, and BL region ($\Gamma_1$). The sum of the angle of attack of the wing and the angle of attack induced by the gust does not drop below zero during the gust entry period. The wake shed by the trailing-edge of the wing can only attenuate the circulation in the $\Gamma_1$ region but cannot reverse its sign. The only remaining factor that can reverse the sign of the circulation is the pitching of the wing. The pitch down action of the wing imparts a physical rotation or circulation in the flow that opposes the circulation direction of the LEV, creating a circulatory force contribution opposite to that of the LEV. This effect has been referred to as the `Magnus' component of the circulatory contribution to the force~\cite{stevens2017experiments} or the virtual camber induced by pitching~\cite{leishman2006principles}. The presented circulation trends obtained from time-resolved PIV demonstrate that the pitching action of the wing produces physical circulation that resides in the combined boundary layer and secondary vorticity region.

\section{Conclusion}
This work demonstrated experimentally that simple proportional closed-loop feedback control based on pitch acceleration $\ddot{\alpha}$ input can successfully mitigate large-amplitude transverse gust encounters without \textit{a priori} knowledge of gust strength or onset time. It was also shown to successfully achieve its objective for various gust strengths, upwards and downwards gusts, and pre- and post-stall angles of attack. Additionally, it succeeded at mitigating lift transients experienced by the wing after exiting the gust (gust-recovery region).

A dynamical systems treatment of the classical aerodynamic model of Wagner revealed several physical features important to vehicle control when wing pitch is used as an input. The effect of pitch location, pitch input choice, and chosen gain along with the physical mechanisms responsible are summarized in table~\ref{tab:my-table}. Pitching aft the midchord is found to create an unstable feedback controller because a portion of the added-mass force opposes the lift control goal. Instantaneous added-mass and circulatory dynamics are found to feed-through directly to lift output, and thus may amplify sensor noise. Choosing a higher-order control input, such as $\ddot{\alpha}$, mitigates this issue. Additionally, it was found that a maneuvering wing control strategy heavily relies on unsteady flow physics, such as added-mass and rotational circulation, that arise from quick maneuvers. 

Time-resolved flowfield and force measurements acquired at various gust ratios and starting angles of attack $\alpha^\text{ref.}$ revealed salient flow physics that illustrate how closed-loop actuation mitigates the gust within the main gust region as well as the gust recovery region (region where the wing is outside the gust but continues to experience lift transients). As the wing enters an upward or downward transverse gust, it pitches such that it reduces the changes in the total effective angle of attack. The pitch motion also creates added-mass forces that counteract the non-circulatory gust contribution to lift. In the gust recovery region, the uncontrolled wing at a high reference angle of attack experiences a lift dip due to the formation of a trailing-edge vortex, followed by a lift peak due to the formation of a leading-edge vortex. The controlled wing mitigated the secondary lift transients by suppressing the formation of the trailing-edge vortex and maintaining a reduced angle of attack during the presence of the leading-edge vortex. Contrary to upward gusts, downward gust encounters at high starting angles of attack do not result in a trailing-edge vortex in the gust recovery region, and thus the lift does not dip below steady state during the gust exit period. The controller is also able to mitigate the secondary lift peak for these encounters by maintaining a lower angle of attack.

Finally, decomposing the lift force and quantifying the circulations in various parts of the flowfield from time-resolved PIV measurements reveals how circulation, added mass, and non-circulatory gust effects affect the forces. Through this decomposition, it is shown that the added-mass force plays an integral role in gust mitigation by counter-balancing the non-circulatory gust force. In addition, the pitching motion is shown to diminish the circulatory contribution to the lift force. While both the controlled and uncontrolled wings develop strong LEVs during the gust entry period, the pitching motion of the controlled wing creates a large amount of circulation in the wing's boundary layer that significantly reduces the circulatory force contribution, reversing its sign. This finding illustrates that it is not sufficient to only examine the LEV to make inferences regarding the circulatory contribution to the lift force in the context of a wing pitching in a gust. It is also shown that the Magnus force or the virtual camber used in unsteady aerodynamic modeling has its basis in physical circulation produced during pitching and residing in the wing's boundary layer and secondary vorticity. 

\section*{Acknowledgments}
The authors gratefully acknowledge support from the Air Force Office of Scientific Research under grant FA9550-16-1-0508 and the National Science Foundation under grant 1553970.  

\bibliography{Refs}

\end{document}